\documentclass[11pt,a4paper]{article}
\pdfoutput=1 
\usepackage{jheppub2}
\usepackage[table]{xcolor}
\usepackage[utf8]{inputenc}
\usepackage{colortbl}
\usepackage{varwidth}
\RequirePackage{ifpdf} 
\usepackage{amsmath} 
\usepackage{mathtools}
\usepackage{pstricks}
\usepackage[final]{pdfpages} 
\usepackage{ifpdf} 
\usepackage{slashed}
\usepackage{natbib}
\setcitestyle{square, comma, numbers,sort&compress}
\usepackage{hyperref}
\usepackage[normalem]{ulem}
\usepackage{color}
\usepackage{soul,xcolor}
\usepackage{xcolor}
\definecolor{urlblue}{rgb}{0.2,0.4,0.7}
\definecolor{citegreen}{rgb}{0,0.6,0.2}
\definecolor{linkred}{rgb}{0.9,0.2,0.1}
\setstcolor{red}
\usepackage{hyperref}
\hypersetup{
	colorlinks=true, citecolor=citegreen, linkcolor=blue, urlcolor=urlblue}
\usepackage{float}
\usepackage[caption = false]{subfig}
\usepackage{graphics}
\usepackage{etoolbox} 
\usepackage{fixmath}
\usepackage{psfrag}
\usepackage[utf8]{inputenc} 
\usepackage[T1]{fontenc}

\usepackage{amsfonts}
\usepackage{autobreak}
\usepackage{marginnote}
\usepackage{enumitem}
\usepackage{appendix}

\newcommand{\NOdisplay}[1]{ }

\title{Two-loop form factors for Dark Matter annihilation to colored Standard Model particles}

\author{Warsimakram Katapur$^{a}$, Ambresh Shivaji$^{a}$}
\emailAdd{ik.wasimakram83@gmail.com,ashivaji@iisermohali.ac.in}
\affiliation{
$^a$Indian Institute of Science Education and Research Mohali,\\
Knowledge City, Sector 81, SAS Nagar, Manauli, Punjab 140306, India}

\preprint{}

\abstract{We investigate a UV-complete model in which the dark matter (DM) particle interacts with gluons through a colored scalar mediator. This framework provides a phenomenologically viable scenario testable at hadron colliders. While mono-jet signatures are relevant for collider searches, zero-jet processes correspond to complete annihilation of Standard Model (SM) particles into DM, contributing to the relic density. In this work, we study dark matter annihilation into SM colored particles, which in our model arises at leading order from loop-induced processes. We compute the relevant two-loop QCD amplitudes for both gluon and quark channels in dark matter production or annihilation. The amplitudes are decomposed into scalar form factors using the projector technique. Using integration-by-parts (IBP) identities, we obtain analytical expressions for the form factors in terms of master integrals. Ultraviolet divergences are removed via counterterm renormalization, yielding UV-finite results. These results would enable predictions for dark matter production or annihilation into SM colored particles at next-to-leading order (NLO) in QCD.}

\begin{document}
\allowdisplaybreaks[4]
\unitlength1cm
\maketitle
\flushbottom

\section{Introduction}
\label{sec:intro}

The search for dark matter (DM) is a major undertaking in the field of particle physics. A number of experiments, representing a diverse range of approaches to the dark matter problem, each with its own strengths and limitations, are currently collecting data ~\citep{XENON:2023cxc,Fermi-LAT:2015att,ATLAS:2021kxv,CMS:2013cmj}. Assuming a particle candidate exists for DM, it could be investigated through direct, indirect and  collider searches. Direct searches, such as in underground experiments, are based on the scattering of dark matter with standard model (SM) particles, DM+SM $\rightarrow$ DM+SM~\citep{Hicyilmaz:2021}. Indirect searches look for an excess of events over astrophysical backgrounds from dark matter annihilation into SM particles, DM+DM $\rightarrow$ SM+SM~\citep{Goodenough:2009gk}. At colliders, dark matter can be produced via SM+SM $\rightarrow$ DM+DM scattering~\citep{Askew:2014kqa}.

Simplified models have become a favored theoretical tool, capturing key collider and cosmological phenomenology while maintaining a transparent, minimal parameter space~\citep{Abdallah:2015ter,DeSimone:2016fbz,Morgante:2018tiq}.
In these models, DM interactions with SM fields are mediated by new particles whose properties can be systematically studied at colliders, in direct detection experiments, and through indirect searches.

In the simplified model approach, both the DM and the mediator can be a scalar, vector, or fermion. The mediators may be charged under the SM gauge groups $SU(3)_C$ or $SU(2)_L \times U(1)_Y$.
If the mediators carry color charge, they can be produced efficiently at hadron colliders through QCD processes, offering strong discovery prospects. Dark matter models with colored mediators have been extensively studied in Refs.~\citep{Goodman:2011jq,DiFranzo:2013vra,An:2013xka,Godbole:2015gma,Ko:2016zxg,Godbole:2016mzr,Arina:2020udz,Arina:2023msd}.
Colored scalar mediators, in particular, are motivated by UV-complete theories, including supersymmetry~\citep{Martin:2007hn}.

From the collider standpoint, colored mediators enable rich signals such as mono-jet plus missing energy  and multi-jet plus missing energy. Cosmologically, these mediators also drive dark matter annihilation into colored SM particles, thereby influencing the dark matter thermal relic density. 
The presence of colored mediators makes the observables sensitive to QCD radiative corrections~\citep{Arina:2023msd,Bringmann:2015cpa,Borschensky:2018zmq}. 
In several theoretical frameworks, dark matter annihilation and production proceed predominantly via loop-induced processes, emphasizing the need for higher-order QCD corrections~\citep{Godbole:2015gma,Mattelaer:2015haa,Godbole:2016mzr}.

In this paper, we study a simplified scalar dark matter model in which DM interacts primarily with gluons through a colored scalar mediator. In such a framework, DM annihilation and production occur via loop-induced processes~\citep{Godbole:2016mzr}. Our goal is to compute the two-loop amplitudes for DM annihilation into quarks and gluons, which constitute the essential ingredients for obtaining next-to-leading order (NLO) QCD predictions of the annihilation cross section.

The computation of two-loop amplitudes in Beyond the Standard Model (BSM) scenarios has become increasingly important in recent years, driven by the precision goals of both collider and cosmological studies~\citep{Braathen:2019zoh,McKay:2017xlc,Klasen:2016qyz}. Two-loop amplitude calculations provide the necessary ingredients to capture both ultraviolet (UV) and infrared (IR) structures consistently and to match with effective field theory (EFT) descriptions in appropriate limits~\citep{Slavich:2020zjv,Bahl:2020jaq}. Techniques such as form-factor decomposition, integration-by-parts (IBP) reduction, and the identification of master integrals have enabled analytic control over multi-loop computations even in BSM frameworks~\citep{Anastasiou:2006hc,Hessenberger:2016atw}. These advances make it possible to achieve precision comparable to Standard Model two-loop studies such as $H \to gg$ and $H \to q {\bar q}$, extending predictive power to a broad class of BSM scenarios including the dark matter models.

The rest of the paper is organized as follows: In section~\ref{sec:dmannihilation}, we briefly review the model and discuss relevant dark matter annihilation channels. The form factor calculations for annihilation channels are presented in section~\ref{sec:FF}. The details of UV renormalization of form factor and the subtraction of universal IR structure are given in section~\ref{sec:renorm}. Various checks made on the form factors are described in section~\ref{sec:checks}. Finally, in section~\ref{sec:results}, the form factors are given as an expansion in $\epsilon$ parameter of dimensional regularization. We conclude in section~\ref{sec:conclusions}.

\section{Dark Matter annihilation to quarks and gluons}
\label{sec:dmannihilation}
\begin{figure}[H]
    \centering
    \begin{minipage}{0.25\textwidth} 
        \centering
        \includegraphics[width=\textwidth]{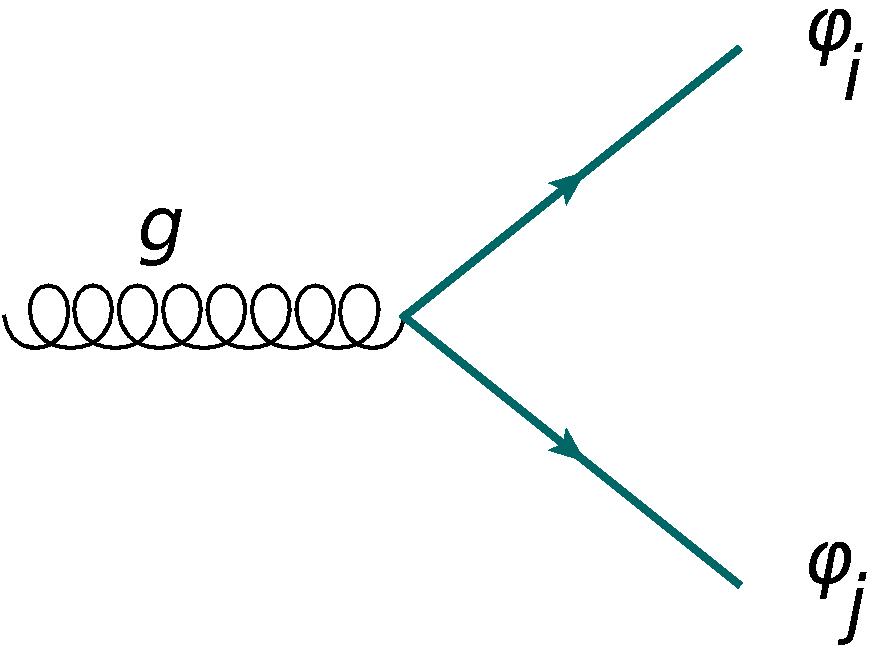}
    \end{minipage}%
    \hspace{0.2cm}
    \begin{minipage}{0.2\textwidth} 
        \centering
        \begin{equation*}
            i g_s T^a_{ij} (p_2^\mu - p_3^\mu)
        \end{equation*}
    \end{minipage}    
    \vspace{0.5cm}
    \begin{minipage}{0.2\textwidth}
        \centering
        \includegraphics[width=\textwidth]{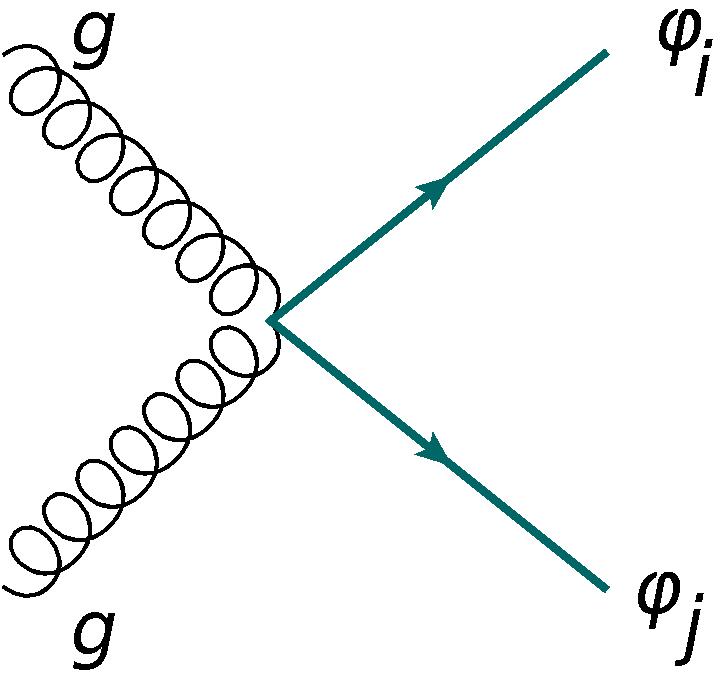}
    \end{minipage}%
    \hspace{0.2cm}
    \begin{minipage}{0.2\textwidth}
        \centering
        \begin{equation*}
            i g_s^2 g_{\mu \nu} (T^a_{jk} T^b_{ki} + T^b_{jk} T^a_{ki})
        \end{equation*}
    \end{minipage}
    \vspace{0.5cm}
    \begin{minipage}{0.25\textwidth}
        \centering
        \includegraphics[width=0.8\textwidth]{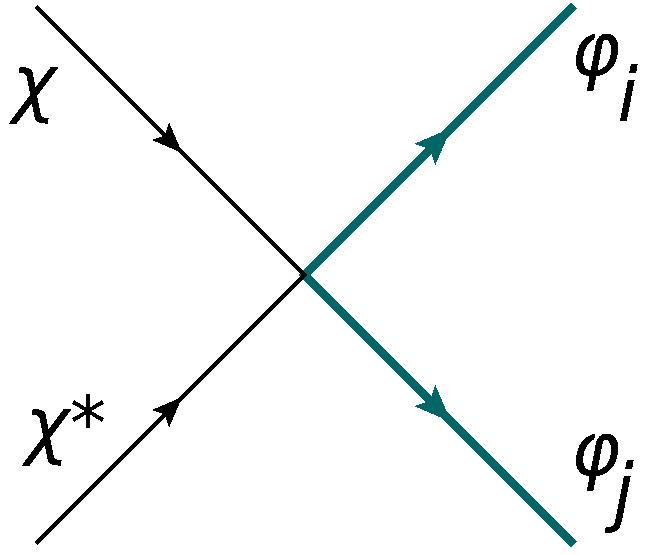}
    \end{minipage}%
    \hspace{0.1cm}
    \begin{minipage}{0.2\textwidth}
        \centering
        \begin{equation*}
            i \lambda_d \delta^{ij}
        \end{equation*}
    \end{minipage}    
    \caption{Feynman rules for interaction vertices involving dark matter $\chi$, colored mediator $\phi$ and gluons $g$.}
    \label{fig:vertices}
\end{figure}

In the minimal version of the dark matter model with a colored scalar mediator, the interaction of the mediator $\phi$ with gluons is governed by the strong coupling constant $g_s$, while its coupling with the dark matter candidate $\chi$ is governed by the parameter $\lambda_d$.
The model Lagrangian is given by Ref.~\citep{Godbole:2015gma}
\begin{equation}
 {\mathcal{L}} = {\mathcal{L}}_{\rm SM} +  \partial_{\mu}\chi^{*} \partial^{\mu}\chi- m_{\chi}^2 |\chi|^2 + (D_{\mu}\phi)^{\dagger} D^{\mu}\phi - m_{\phi}^2|\phi|^2+ \lambda_{d}\chi^{*} \chi \phi^* \phi, 
\end{equation}
where $m_\chi$ is the mass of the complex scalar dark matter $\chi$, $m_\phi$ is the mass of the colored complex scalar mediator $\phi$ and $D_{\mu}\phi= \partial_{\mu}\phi - i g_s \dfrac{\lambda^a}{2} G^a_{\mu} \phi$. The new interaction vertices in this model are shown in Fig.~\ref{fig:vertices}.
Under $SU(3)_C$, there is freedom in choosing the representation for the mediator particle $\phi$. We take $\phi$ in fundamental representation. In the large $m_\phi$ limit, the dark matter $\chi$ develops an effective interaction with the gluons which is governed by the following gauge invariant Lagrangian~\citep{Goodman:2010ku}:
\begin{equation}
\mathcal{L}_{\text{eff}} \equiv \dfrac{\alpha_s}{m_\phi^2} |\chi|^2 G^{\mu\nu, a} G^a_{\mu\nu},
\end{equation}
where \[
G_{\mu\nu}^a = \partial_\mu G_\nu^a - \partial_\nu G_\mu^a + g_s f^{abc} G_\mu^b G_\nu^c
\].

\begin{figure}[h]
    \centering
    \begin{minipage}{0.6\textwidth}
        \centering
        \includegraphics[width=9cm, height=2cm]{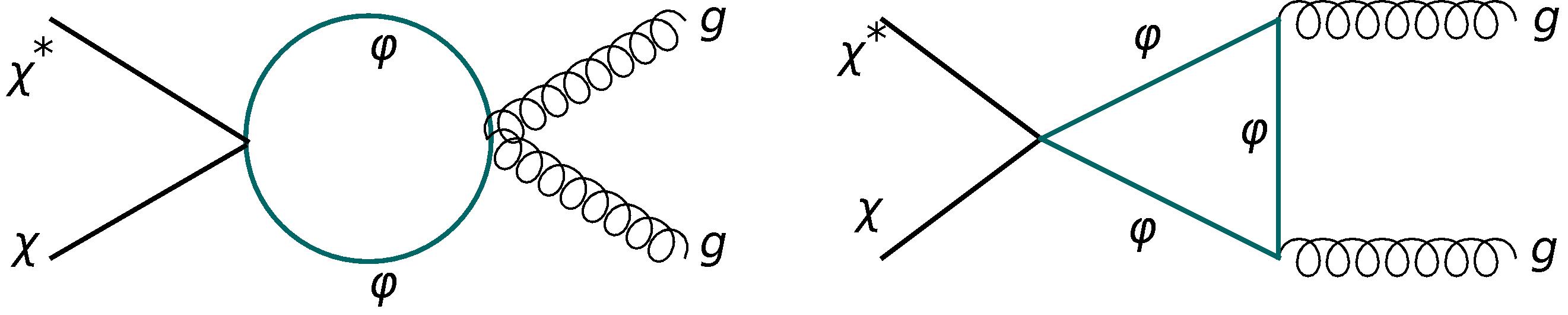}
    \end{minipage}%
    \caption{Leading order Feynman diagrams for \(\chi \chi^* \rightarrow  g g \).}
    \label{fig:ggDM}
\end{figure}

\begin{figure}[h]
    \centering
    \begin{minipage}{0.6\textwidth}
        \centering
        \includegraphics[width=9cm, height=3cm]{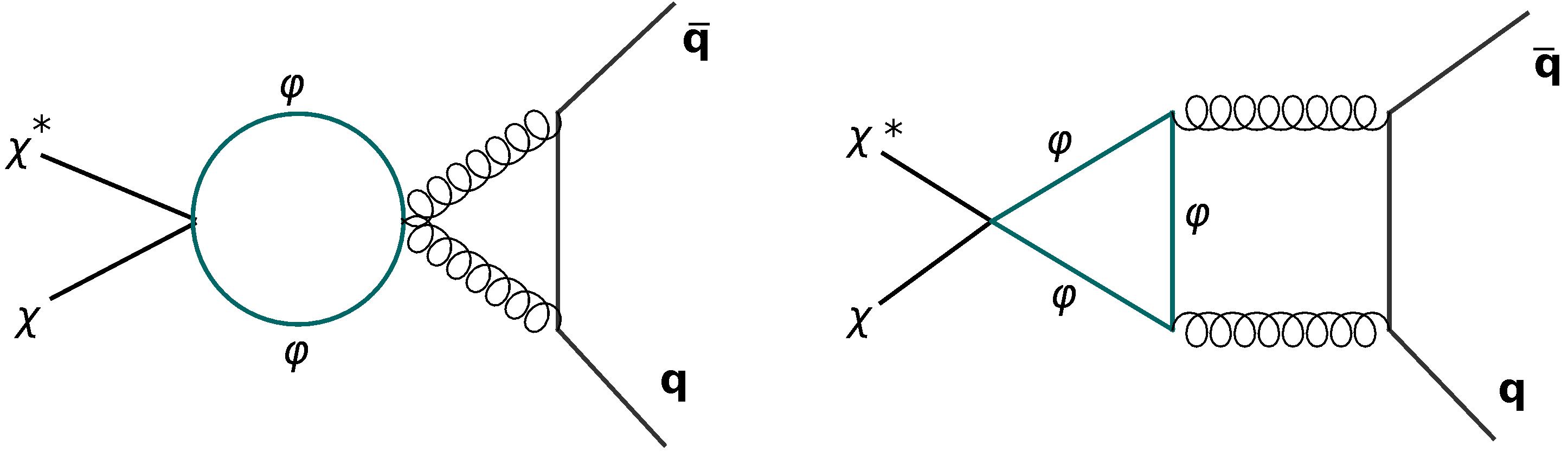}
    \end{minipage}%
    \caption{Leading order Feynman diagrams for  $\chi \chi^* \to q \bar{q}$.}
    \label{fig:qqDM}
\end{figure}

The parton-level contributions to the annihilation of dark matter pairs into Standard Model particles arise from the channels
$\chi \chi^{*} \to  g g$ and 
 $\chi \chi^{*} \to q \bar{q}$, where $g$ denotes a gluon and $q$ represents a quark. Since the dark matter candidate does not directly couple to the quarks or gluons, these annihilation channels are loop-induced processes. The leading order contribution to $\chi \chi^* \to g g$ comes from the one-loop diagrams shown in Fig.~\ref{fig:ggDM}, while 
$\chi \chi^* \to q {\bar q}$ receives its leading order contribution from the two-loop diagrams illustrated in Fig.~\ref{fig:qqDM}. Because of the four-point contact interaction between $\phi$ and $\chi$, the annihilation processes effectively correspond to $1 \to 2$ transitions. From this perspective, the triangle diagram of Fig.~\ref{fig:ggDM} is analogous to the one-loop amplitude for $H \to g g$ with colored quarks running in the loop. Similarly, the box-triangle topology of the $\chi \chi^* \to q {\bar q}$ process, shown in Fig.~\ref{fig:qqDM}, resembles the two-loop QCD corrections to $H \to b {\bar b}$.

{In the present work, we aim to compute the two-loop contribution to the $\chi \chi^* \to g g$ amplitude. The leading-order one-loop result is already known~\citep{Godbole:2015gma}. The gluon-channel diagrams are analogous to the squark-loop contributions in $gg \to H$ processes in supersymmetric theories. However, unlike the coupling of the colored scalar with the dark matter pair, the coupling of a squark with the Higgs boson is mass-dependent. This difference affects the renormalization of the amplitude in the two cases. The result for the two-loop amplitude for $gg \to H$ via squarks in the loop is presented in Ref.~\citep{Anastasiou:2006hc}. Similar to the gluon channel, the two-loop diagrams for the quark channel also arise in supersymmetric theories; however, no such result is currently available in the literature.}

\section{Form factors for annihilation channels}
\label{sec:FF}

The annihilation rates of dark matter into gluon and quark channels can be conveniently organized in term of the form factors. 

\subsection{Form factor for $gg$ channel}
\label{sec:ggchannel}
The amplitude for the annihilation process, $\chi\chi^* \to g(p_1) g(p_2)$ can be written as,
\begin{eqnarray} 
    {\cal A} (\chi\chi^* \to gg) = \mathcal{M}^{\mu \nu} \epsilon_\mu(p_1) \epsilon_\nu(p_2),
\end{eqnarray}
where, $\epsilon_\mu(p_1)$ and $\epsilon_\nu(p_2)$ are polarization vectors of the gluons.  
Similar to the $H \to gg$ process, the amplitude can be expressed in terms of Lorentz-invariant form factors as
\begin{equation}
  \mathcal{M}^{\mu \nu}= \delta ^{ab} \Big( \dfrac{s}{2} {\cal F}_1~ \ g^{\mu \nu} + {\cal F}_2~  p_1^\nu p_2^\mu  \ \Big),
\end{equation}
where $a$ and $b$ denote the color indices of the gluons, and $s=(p_1+p_2)^2$. The two form factors ${\cal F}_1$ and ${\cal F}_2$ are related by current conservation,
\begin{equation}
    p_{1,\mu} \mathcal{M}^{\mu \nu} = p_{2,\nu} \mathcal{M}^{\mu \nu} = 0,
\end{equation}
which implies ${\cal F}_2 = - {\cal F}_1 $. Hence, the amplitude depends on a single independent form factor ${\cal F}_1={\cal F}_{gg}$, and can be written as 
\begin{equation}
  \mathcal{M}^{\mu \nu}= {\cal F}_{gg}~ \delta ^{ab} \Big( \dfrac{s}{2}  g^{\mu \nu} -   p_1^\nu p_2^\mu  \ \Big)  .
\end{equation}

The form factor ${\cal F}_{gg}$ can be expanded perturbatively in the strong coupling $\alpha_s$. It can be extracted from the amplitude using a suitable projector ${\cal P}_{\mu\nu}$ defined by, 
\begin{equation}
    \mathcal{P}^{\mu \nu}\mathcal{M}_{\mu \nu}= {\cal F}_{gg}.
\end{equation}
Choosing a projector with the same tensor structure as the amplitude, and working in $d$ dimensions, we obtain
\begin{equation}
    \mathcal{P}^{\mu \nu}=  \dfrac{1}{(d-2)(s/2)} \Bigg(g^{\mu \nu} -\dfrac{p_1^\nu p_2^\mu }{(s/2)}  \Bigg).
\end{equation}

The one-loop leading order result for the form factor is
\begin{equation}
{\cal F}^{\rm 1L}_{gg} = \dfrac{2i\alpha_s \lambda_d}{4\pi s^2} \left( s + m_{\phi}^2 \ \log \left[ \dfrac{2 m_{\phi}^2 - s + \sqrt{s (s - 4 m_{\phi}^2)}}{2 m_{\phi}^2} \right]^2 \right) .
\label{eq:Amp1Lgg}
\end{equation}
The individual bubble and triangle diagrams contributing to this process contain ultraviolet (UV) divergences, but the total amplitude is UV finite, as expected. No infrared (IR) divergences occur in either the massive or massless cases at the level of individual diagrams.

\begin{figure}[H]
 \centering
 \includegraphics[width=14cm, height=2cm]{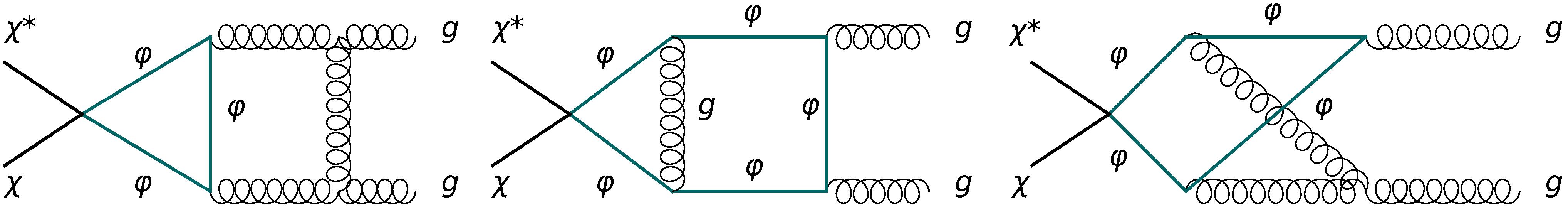}
 \caption{Topmost topologies in $\chi \chi* \rightarrow g g$ , black lines are massless gluons, green lines are massive mediators and thin black lines are dark matter particles. }
 \label{fig:gg2LTopDiagrams}
\end{figure}

In this work, we focus the two-loop expression of the form factor $\mathcal{F}_{gg}$ at $\mathcal{O}(\alpha_s^2)$. 
The two-loop Feynman diagrams containing the maximum number of propagators that contribute at this order are shown in 
Figure~\ref{fig:gg2LTopDiagrams}. 
We introduce a dimensionless ratio between the kinematic variable $s$ and the scalar mass $m_\phi$ as,
\begin{equation}
  \frac{s}{m_\phi^2} = -\,\frac{(1 - x)^2}{x}.
  \label{eq:ggParametrization}
\end{equation}
This parametrization simplifies the analytic structure of the integrals by rationalizing the square roots that arise when the system of differential equations is brought to its canonical form.

The relevant Feynman diagrams are generated using {\tt QGRAF}~\citep{Nogueira:1991ex}, while the algebraic manipulations are carried out in {\tt FORM}~\citep{Vermaseren:2000nd} and {\tt Mathematica}. After applying the projector, the tensor amplitudes reduce to scalar integrals.  We use the program {\tt Reduze2}\citep{vonManteuffel:2012np} to perform momentum shifts and external leg exchanges, allowing the integrals to be mapped into a minimal number of independent integral families.
For our case, this results in three distinct integral families, each containing seven propagators, referred to as PL1, PL2, and NP. These are listed in Table.~\ref{table:ggIntegralFamilies}.
\begin{table}[H]
    \centering
    \begin{tabular}{|c|c|c|}
        \hline
        \textbf{PL1} & \textbf{PL2} & \textbf{NP} \\
        \hline
        $\{k_1,0\}$ & $\{k_1, m_\phi\}$ & $\{k_1, m_\phi\}$ \\
        \hline
        $\{k_1-k_2,m_\phi\}$ & $\{k_1-k_2,0\}$ & $\{k_1-k_2,0\}$ \\
        \hline
        $\{k_2,m_\phi\}$ & $\{k_2, m_\phi\}$ & $\{k_1-k_2-p_1,0\}$ \\
        \hline
        $\{k_1+p_1+p_2,0\}$ & $\{k_2+p_2, m_\phi\}$ & $\{k_2+p_1, m_\phi\}$ \\
        \hline
        $\{k_2+p_1+p_2,m_\phi\}$ & $\{k_1+p_1+p_2, m_\phi\}$ & $\{k_1+p_1+p_2, m_\phi\}$ \\
        \hline
        $\{k_1+p_1,0\}$ & $\{k_2+p_1+p_2, m_\phi\}$ & $\{k_2+p_1+p_2, m_\phi\}$ \\
        \hline
        $\{k_2+p_1, m_\phi\}$ & $\{k_1+p_2, m_\phi\}$ & $\{k_1+p_1,0\}$ \\
        \hline
    \end{tabular}
    \caption{Integral families for PL1, PL2, and NP.}
    \label{table:ggIntegralFamilies}
\end{table}

The gluon channel integral family is defined by,
\begin{equation}
    I_{gg}(a_1, a_2, \dots, a_n) =  \left( m_{\phi}^{2\epsilon} e^{\gamma_E \epsilon}\right)^2 
    \int \int \dfrac{d^d k_1}{(i \pi)^{d/2}} \dfrac{d^d k_2}{(i \pi)^{d/2}} \dfrac{1}{P_1^{\nu_1} \cdots P_7^{\nu_7}}.
\end{equation}
Here, $k_1$ and $k_2$ denote the loop momenta, $\gamma_E$ is the Euler-Mascheroni constant and $\{\nu _i\} $ = $\{\nu_1 \nu_2 \nu_3 \nu_4 \nu_5 \nu_6 \nu_7\}$ are powers of the inverse propagators $P_i$.

Only a single Mandelstam variable,
$s$, appears in this case. Since $s$ is invariant under the exchange of external legs, the scalar integrals remain unaffected by such mappings. The two-loop scalar integrals in the form factor are reduced to a minimal basis of master integrals (MIs) using integration-by-parts (IBP)~\citep{Tkachov:1981wb,Chetyrkin:1981qh} and Lorentz-invariance (LI) \citep{Gehrmann:1999as} identities. We make use of programs such as \texttt{Kira3.0}~\citep{Lange:2025fba}, \texttt{LiteRed}~\citep{Lee:2012cn,Lee_2014} combined with \texttt{Mint}~\citep{Lee:2013hzt} and \texttt{FIRE}~\citep{A.V.Smirnov_2008,SMIRNOV20132820,SMIRNOV2015182} for these reductions. A few integrals were mapped to MIs available in literature using \texttt{Azurite}~\citep{Georgoudis:2016wff}. For computations involving polylogarithms functions-such as expansions and numerical evaluations, we employ {\tt PolyLogTools}~\citep{Duhr:2019tlz}. The MIs obtained from \texttt{FIRE} for the planar and non-planar families are as follows.

\paragraph{Planar family PL1:}
\begin{equation}
\begin{aligned}
\{ &\text{PL1}[0,1,0,0,0,0,1],\,
\text{PL1}[0,1,1,1,0,0,0],\,
\text{PL1}[0,2,1,0,1,0,1],\,
\text{PL1}[0,2,1,0,1,1,0],\\
&\text{PL1}[0,2,1,2,0,0,0],\,
\text{PL1}[0,2,2,0,0,0,0],\,
\text{PL1}[0,2,2,0,1,0,0],\,
\text{PL1}[0,2,2,0,1,1,0],\\
&\text{PL1}[0,2,2,1,0,0,0],\,
\text{PL1}[0,3,1,0,1,1,0],\,
\text{PL1}[1,0,1,1,0,0,0],\,
\text{PL1}[1,1,0,0,1,0,1],\\
&\text{PL1}[1,1,1,1,1,0,0],\,
\text{PL1}[1,2,0,0,1,0,0],\,
\text{PL1}[1,2,0,2,0,0,0],\,
\text{PL1}[2,0,2,1,1,0,0] \}.
\end{aligned}
\end{equation}

\paragraph{Planar family PL2:}
\begin{equation}
\begin{aligned}
\{ &\text{PL2}[0,0,1,0,0,0,1],\,
\text{PL2}[0,1,1,0,1,0,0],\,
\text{PL2}[0,1,1,0,1,0,1],\,
\text{PL2}[0,1,1,1,1,0,1],\\
&\text{PL2}[0,1,2,0,1,0,0],\,
\text{PL2}[1,0,1,0,1,0,0],\,
\text{PL2}[1,0,1,0,1,0,1],\,
\text{PL2}[1,1,0,1,1,0,0],\\
&\text{PL2}[1,1,0,1,1,2,0],\,
\text{PL2}[1,1,0,1,2,0,0],\,
\text{PL2}[1,1,0,2,1,0,0],\,
\text{PL2}[1,1,1,1,1,0,0],\\
&\text{PL2}[1,1,1,1,2,0,0],\,
\text{PL2}[1,2,0,0,0,1,0],\,
\text{PL2}[2,0,1,1,1,1,0],\,
\text{PL2}[2,0,2,0,1,1,0] \}.
\end{aligned}
\end{equation}

\paragraph{Non-planar family NP:}
\begin{equation}
\begin{aligned}
\{ &\text{NP}[0,0,1,1,1,0,0],\,
\text{NP}[0,0,1,2,1,0,0],\,
\text{NP}[1,0,0,1,0,0,0],\,
\text{NP}[1,0,0,1,1,0,0],\\
&\text{NP}[1,0,1,1,1,2,0],\,
\text{NP}[1,1,0,1,0,1,0],\,
\text{NP}[1,1,0,1,1,0,0],\,
\text{NP}[1,1,0,1,1,1,0],\\
&\text{NP}[1,1,0,1,2,0,0],\,
\text{NP}[1,1,0,2,1,0,0],\,
\text{NP}[1,1,1,1,1,1,0],\,
\text{NP}[1,2,0,0,0,1,0],\\
&\text{NP}[2,1,0,1,1,1,0] \}.
\end{aligned}
\end{equation}

All of the above integrals can be mapped to the 18 master integrals known in the literature. Analytical results for these master integrals are available in Ref.~\citep{Anastasiou:2020qzk}.

\subsection{Form factor for $ q \bar{q } $ channel}
\label{sec:qqchannel}
The amplitude for the annihilation process, $ \chi \chi*  \rightarrow q(p_1) \bar{q}(p_2) $
can be expressed as,
\begin{equation}
    \mathcal{A}({\chi \chi*\rightarrow q {\bar q}  })= \bar{u}(p_1)~ \delta^{ij} {\cal F}_{qq}~\ v(p_2)
\end{equation}
where, $i$ and $j$ are color indices of quarks and  ${\cal F}_{qq}$ is the form factor which as in the $gg$ case can be computed perturbatively in $\alpha_s$.  
Here $p_1^2=p_2^2= m_q^2$ and $p_1.p_2=\dfrac{s}{2}-m_q^2$. 
We choose an ansatz for the projector, $$ {\cal P}= \beta~ \bar{v}(p_2)~u(p_1) $$
where $\beta$ is a scalar constant, such that it satisfies the projector condition,
\begin{equation}
\sum_{\text{pol.}} {\cal P} \, \mathcal{A} = {\cal F}_{qq}.
\end{equation}
Solving the above equation gives, 
\begin{eqnarray}
    \beta = \dfrac{1}{\text{ Tr}[\slashed{p}_1 \slashed{p}_2- m_q^2 ]} = \dfrac{1}{d \ (p_1.p_2- m_q^2)}
\end{eqnarray}


This results in a projector,
\begin{equation}
    {\cal P}= \dfrac{1}{d \ (p_1.p_2- m_q^2)} \  \bar{v}(p_2)~ u(p_1).
\end{equation}

At one-loop, the process $ \chi \chi* \to q\bar {q } $ receives no contributions from quark channel in the Gluphilic scalar dark matter model. The representative Feynman diagrams at two-loop order due to the allowed vertices in the model are shown in Figure~\ref{fig:qqDM}.
The external quarks we consider can be massive or massless. For massless quarks, the contribution is always zero since the quark leg with two vertices and a quark propagator leads to a trace over an odd number of $\gamma$ matrices. But when massive quarks such as $b$ or $t$ quarks are considered, there is a non-zero contribution due to the trace over an even number of gamma matrices. Hence, it is necessary to consider massive external quarks when evaluating dark-matter pair annihilation or production via the quark channel. This is expected to be true to all loop orders due to chiral symmetry. 
This also holds for $H \to q {\bar q}$ process. For example, when only the $Z$ boson is considered in the loop, the massless quarks contribute zero. Whereas, there is a non-zero contribution from massive quarks. This has been verified explicitly using {\tt MadGraph5\_aMC@NLO}~\citep{Alwall:2014hca}. This observation affects constraints on the annihilation rates of the dark matter pair to a quark pair, which can only be massive.

At two-loops, there are reducible and irreducible diagrams. The reducible diagrams contribution is zero since they have a gluon propagator connecting the $\phi$ loop and gluon loops, and their color factor vanishes due to a trace over a single Gell-Mann matrix. 

For the quark channel at two-loop order, there are 20 master integrals (MI) for the massive mediator, and only 1 planar topology is sufficient for mapping all the Feynman diagrams,
 which is shown in Table.~\ref{table:PL1qq}. 

\begin{table}[H]
    \centering
    \begin{tabular}{|c|}
        \hline
        \textbf{PL1} \\
        \hline
        $\{k_1, m_q\}$ \\
        \hline
        $\{k_2, m_\phi\}$ \\
        \hline
        $\{k_1 + p_1, 0\}$ \\
        \hline
        $\{k_1 - k_2 + p_1, m_\phi\}$ \\
        \hline
        $\{k_1 - p_2, 0\}$ \\
        \hline
        $\{k_2 - p_1 - p_2, m_\phi\}$ \\
        \hline
        $\{k_2 + p_1, 0\}$ \\
        \hline
    \end{tabular}
    \caption{Integral family for PL1 in the quark channel.}
    \label{table:PL1qq}
\end{table}

The quark channel integral family is defined by,
\begin{equation}
\mathcal{I}_{qq}(n_1, \ldots, n_7) = 
\left( \dfrac{1}{\Gamma(1 + \epsilon)} \right)^2
\left( \dfrac{\mu^2}{m_q^2} \right)^{2\epsilon}
\int \dfrac{\mathrm{d}^d k_1}{(i\pi)^{d/2}} \dfrac{\mathrm{d}^d k_2}{(i\pi)^{d/2}} 
\dfrac{1}{D_1^{n_1} \cdots D_7^{n_7}}.
\end{equation}
The 20 master integrals are similar to the master integral available in Ref.~\citep{Mondini:2020uyy} for $H \to c\bar{c}$ with b quark in the loop.

Since in the quark channel, three mass scales $s, m_\phi$ and $m_q$ are involved, the master integrals are expressed in terms of two scaleless variables $w$ and $z$ which are given by, 
\begin{equation}
    -\dfrac{s}{m_\phi^2}=\dfrac{(1-w^2)^2}{w^2}, \ \dfrac{m_q^2}{m_\phi^2}=\dfrac{(1-w^2)^2 z^2}{(1-z^2)^2 w^2}.
\end{equation}

\section{UV renormalization and IR Subtraction}
\label{sec:renorm}
The renormalized amplitude is related to the bare amplitude as,  
\begin{equation}
    \mathcal{A}(\alpha_s, m_\phi, \lambda_d, \mu) = Z_g \mathcal{A}^{0}(\alpha_s^{0}, m_\phi^{0}, \lambda_d^{0})
    \label{eqn:ggrenorm}
\end{equation}
where $Z_g$ denotes gluon wave function renormalization constant. 

The bare amplitude for  dark matter pair annihilation process, $\chi \chi*  \rightarrow g(p_1) g(p_2)$
can be expressed as, 
\begin{equation}
\begin{aligned}
    \mathcal{A}^{0}({\chi \chi^* \rightarrow g g }) &= \mathcal{M}^{\mu\nu}\epsilon_{\mu}(p_1) \epsilon_{\nu}(p_2) \\
    &= \dfrac{i\alpha_s^0 \lambda_d^0 S_{\epsilon} \mu^{-2\epsilon}}{4 \pi} 
    \left(-\dfrac{s}{\mu^2}\right)^{-\epsilon} 
    \delta^{ab} 
    \Bigg[ 
        \dfrac{s}{2} \, (\epsilon_1 \cdot \epsilon_2) 
        - (\epsilon_1 \cdot p_2)(\epsilon_2 \cdot p_1) 
    \Bigg] \\
    &\quad \times 
    \Bigg( 
        \mathcal{M}^{\rm 1L,0}_{gg} 
        + \dfrac{\alpha_s^0 S_{\epsilon} \mu^{-2\epsilon}}{4 \pi} 
        \left(-\dfrac{s}{\mu^2}\right)^{-\epsilon} 
        \mathcal{M}^{\rm 2L,0}_{gg} 
        + \mathcal{O}\Big((\alpha_s^0)^2\Big) 
    \Bigg)
    \label{eqn:gg2L}
\end{aligned}
\end{equation}
where $p_1^2=p_2^2=0$, $p_1.p_2=\dfrac{s}{2}$ and $S_{\epsilon}=(4\pi)^{\epsilon}{\rm exp}(-\epsilon\gamma_E)$. There is a one-to-one correspondence between ${\cal F}_{gg}$ and ${\cal M}_{gg}$.

{In order to remove UV divergences from the matrix elements contributing at $\mathcal{O}(\alpha_s^2)$, we renormalize the strong coupling and the gluon field in $\overline{\text{MS}}$ scheme with $N_f$ = 5 light flavours, whereas the massive colour mediator contributions are renormalized on-shell, at zero momentum. The renormalization process involves evaluating the required counterterm (CT) diagrams and then adding the CT amplitude back to the UV divergent amplitude to obtain finite results after the cancellation of all the divergences.}

We separate \( \mathcal{M}^{\rm 2L,0}_{gg} \) according to,
\begin{equation}
\mathcal{M}^{\rm 2L,0}_{gg}  = M^0_{\rm IR} + M^0_{\rm UV} + \log\left(-\dfrac{s}{\mu^2}\right) M_{\mathrm{fin,scale}}^0 + M_{\mathrm{fin}}^0,
\label{eq:nlobare}
\end{equation}
where the infrared divergences are in $M^0_{\rm IR}$ and the ultraviolet divergences in $M^0_{\rm UV} $. The finite piece $M_{\mathrm{fin,scale}}^0$ contains dependence on renormalization scale $\mu^2$ and $M_{\mathrm{fin}}^0$ corresponds to case $\mu^2=s$.
The infrared divergences follow universal IR structure in $SU(N_c)$ gauge theory at NLO given in Ref.~\citep{Catani:1996vz},
\begin{eqnarray}
    M^0_{\rm IR} &=& -\dfrac{e^{-\gamma_{E} \epsilon}}{\Gamma(1-\epsilon)}\Big(\dfrac{\beta_0}{\epsilon}+\dfrac{2N_c}{\epsilon^2}\Big)\mathcal{M}^{\rm 1L, 0}_{gg}
    \label{eqn:IR}
\end{eqnarray}
where  $\beta_0=\dfrac{11}{3}N_c-\dfrac{2}{3}N_f$ is the one-loop $\beta$ function in SM.

The ultraviolet divergences will have structure as,
\begin{eqnarray}
    M^0_{\rm UV} &=& -\Big(-\dfrac{s}{\mu^2}\Big)^{\epsilon}\Big(\ \delta Z_{\alpha_s} +\delta Z_m (m_\phi^0)^2\dfrac{\partial}{\partial (m^{0}_\phi)^2} + \ \delta Z_{\lambda}  \Big)\mathcal{M}^{\rm 1L, 0}_{gg}
\label{eqn:UV}
\end{eqnarray}
where,

\begin{align}
\delta Z_{\alpha_s} &= Z_{\alpha_s} - 1
= -\left(\frac{\beta_0}{\epsilon} + \frac{T_R}{3\epsilon}\!\left(\frac{\mu^2}{m_\phi^2}\right)^{\!\epsilon}\right), \\[6pt]
\delta Z_{g} &= Z_{g} - 1
= \frac{T_R}{3\epsilon}\!\left(\frac{\mu^2}{m_\phi^2}\right)^{\!\epsilon}, \\[6pt]
\delta Z_{m} &= Z_{m} - 1
= -\frac{3C_F}{\epsilon}, \\[6pt]
\delta Z_{\lambda} &= Z_{\lambda} - 1
= -\frac{3C_F}{\epsilon}.
\end{align}

\begin{figure}[H]
    \centering
    \begin{minipage}{0.4\textwidth} 
        \centering
        \includegraphics[width=\textwidth]{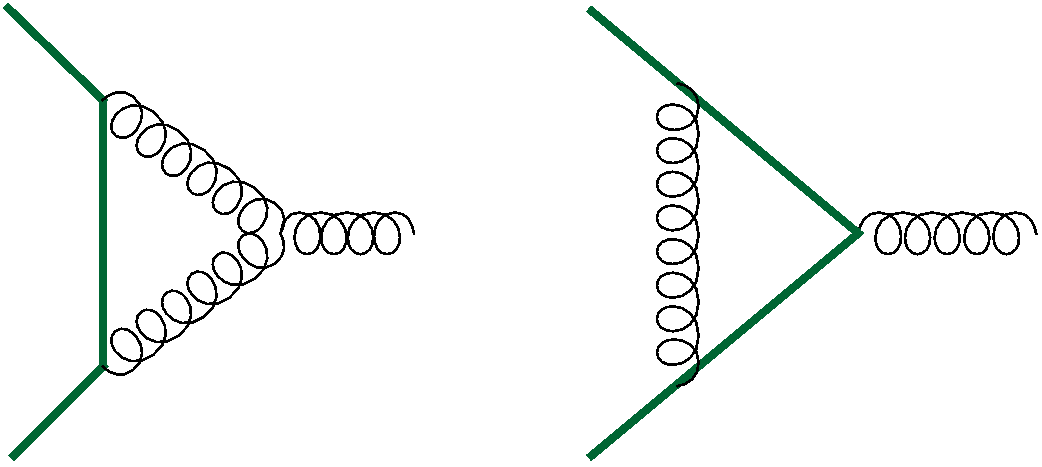}
    \end{minipage}%
    \hspace{1cm}
    \begin{minipage}{0.25\textwidth} 
        \centering
        \begin{equation*}
            \alpha_s^0 = \dfrac{\mu^{2\varepsilon}}{S_\varepsilon} Z_{\alpha_s} \alpha_s\,,
        \end{equation*}
    \end{minipage}    
    \vspace{0.7cm}
    
    \begin{minipage}{0.4\textwidth}
        \centering
        \includegraphics[width=0.75\textwidth]{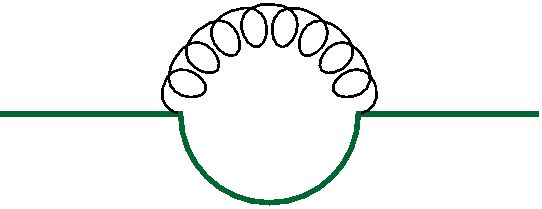}
    \end{minipage}%
    \hspace{1cm}
    \begin{minipage}{0.25\textwidth}
        \centering
        \begin{equation*}
            (m_{\phi}^0)^{2} = Z_m m_{\phi}^2 \, ,
        \end{equation*}
    \end{minipage}
    \vspace{0.7cm}

    \begin{minipage}{0.4\textwidth}
        \centering
        \includegraphics[width=0.5\textwidth]{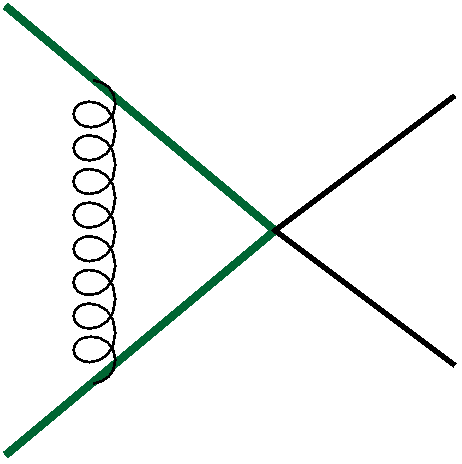}
    \end{minipage}%
    \hspace{1cm}
    \begin{minipage}{0.25\textwidth}
        \centering
        \begin{equation*}
            \lambda_{d}^0 = Z_\lambda \, \lambda_d \,,
        \end{equation*}
    \end{minipage}
    
    \caption{Renormalization of $\alpha_s$ coupling, mediator mass $m_\phi$, and $\lambda_d$ coupling to DM.}
    \label{fig:counterterms}
\end{figure}

Here, $T_R=\dfrac{1}{2}$ and $C_F=\dfrac{N_c^2-1}{2N_c}$. The renormalized couplings and masses are related to the bare couplings and masses as shown in Figure~\ref{fig:counterterms}.

The $m_{\phi}=0$ case is discussed in the Appendix~\ref{app:Appendix}. The renormalization of quark channel is not needed as the two-loop $q\bar{q}$ calculation is a leading order and it is both UV and IR finite.
\section{Checks}
\label{sec:checks}
In order to prove the reliability of our implementation, we have performed several checks for both the $gg$ and $qq$ processes. 

In the gluon channel process, we have verified that the leading order one-loop result vanishes in the large mass limit as expected from the decoupling theorem when couplings are mass independent. The one-loop result is consistent with $gg \to H$ at leading order with squarks in the loop~\citep{Anastasiou:2006hc}. The analytical results for two-loop master integrals are verified through numerical evaluation in {\tt AMFlow}~\citep{Liu:2022chg} to sufficient accuracy. We have reproduced one-loop and two-loop form factors for $g g \to H$ in the standard model which share the same set of master integrals. The results have been cross-checked with Ref.~\cite {Anastasiou:2020qzk} for the bare and renormalized amplitudes. This check provides an important verification of the system of equations relating our master integrals with those available in the literature. The two-loop master integrals can have poles in $\epsilon$ up to order 4. 
However, since the one-loop form factor is finite, we expect only $\dfrac{1}{\epsilon^2}$ and $\dfrac{1}{\epsilon}$ pole structures to appear in the bare amplitude at two-loop. 
The desired cancellation of $\dfrac{1}{\epsilon^4}$ and $\dfrac{1}{\epsilon^3}$ divergences at the level of full amplitude, therefore, serves a very strong check on our calculation.   

After the UV renormalisation, the amplitude has a universal infrared divergent structure which matches the universal IR structure at NLO. To check for gauge invariance, we considered the gluon propagator in $R_\xi$ gauge, and as expected, the form factor is independent of gauge parameter $\xi$.

In the quark channel, since the two-loop form factor is a leading order contribution, we expect it to be free from divergences. Although at the individual diagram level, there are divergences, the divergences get canceled at every order in $\epsilon$ in the form factor. Like in the gluon channel case, the analytical results for master integrals in quark channel are verified through numerical evaluation in {\tt AMFlow} to sufficient accuracy. This also verifies the system of equations relating our master integrals with those available in the literature. The form factor has overall dependence on $m_q$ and remaining 
$m_q$ dependence is regular in the limit $m_q \to 0$ as expected due to chiral symmetry. 
Taking the gluon propagator in $R_\xi$ gauge, the form factor is found to be independent of the gauge parameter $\xi$ ensuring the gauge invariance of the calculation.

\section{Results}
We now present analytical results of form factors for gluon and quark channels when the coloured mediator $\phi$ is treated as massive. As a theoretical cross-check, the massless mediator case is discussed in Appendix~\ref{app:Appendix}. All the results are included as ancillary files.

\label{sec:results}
\subsection{Gluon channel}
The results in the gluon channel will be useful for obtaining NLO QCD predictions for dark matter annihilation cross section or for dark matter production cross section at the LHC. 
At NLO, computation of the interference of leading-order amplitude and the virtual amplitude is required. Since the virtual contribution has divergences up to $\dfrac{1}{\epsilon^2}$, we need leading order contributions up to $\mathcal{O}(\epsilon^2)$. 

In the present work, we have computed leading order contribution up to ${\cal O}(\epsilon^4)$.
The one-loop result for the form factor in terms of the dimensionless variable $x$ defined in Eq.~\ref{eq:ggParametrization} is given by, 
{\allowdisplaybreaks
\begingroup
\begin{eqnarray}
\mathcal{M}^{\rm 1L,0}_{gg}  &=& 
\dfrac{2 x \left((x-1)^2 - x \log^2(x)\right)}{m_\phi^2 (x-1)^4} \nonumber \\
&+&  \dfrac{\epsilon \  2x}{3 m_\phi^2 (x-1)^4} \biggl[
-24 x \mathrm{HPL}(\{-3\},x) + 12 x \log(x) \mathrm{HPL}(\{-2\},x) \nonumber \\
&& + 9 x^2 -6 x^2 \log(x) + 6 x^2 \log(1-x) + 18 x \zeta(3) - 18 x + 2 x \log^3(x) \nonumber \\
&& -3 x \log^2(x) - 6 x \log(1-x) \log^2(x) + \pi^2 x \log(x) + 6 x \log(x) \nonumber \\
&& -12 x \log(1-x) + 6 \log(1-x) + 9 \biggr] \nonumber \\
&+& \dfrac{\epsilon^2 x}{18 m_\phi^2 (x-1)^4} \biggl[
-24 \left(3 x^2 + \pi^2 x + 3 x \log^2(x) - 6 (x + 2 x \log(1-x)) \log(x) - 3\right) \mathrm{HPL}(\{-2\},x) \nonumber \\
&& + 144 x \mathrm{HPL}(\{-2\},x)^2 - 288 x \log(x) \mathrm{HPL}(\{-2,-1\},x) \nonumber \\
&& -144 x (4 \log(1-x) - \log(x) + 2) \mathrm{HPL}(\{-3\},x) + 9 \pi^2 x^2 + 252 x^2 \nonumber \\
&& + 36 x^2 \log^2(x) + 72 x^2 \log^2(1-x) - 216 x^2 \log(x) - 144 x^2 \log(1-x) \log(x) \nonumber \\
&& + 72 x^2 \log(x+1) \log(x) + 216 x^2 \log(1-x) + 216 x \zeta(3) - 72 x \zeta(3) \log(x) \nonumber \\
&& + 432 x \zeta(3) \log(1-x) + \pi^4 x - 6 \pi^2 x - 504 x - 9 x \log^4(x) + 24 x \log^3(x) \nonumber \\
&& + 48 x \log(1-x) \log^3(x) - 72 x \log^2(1-x) \log^2(x) - 9 \pi^2 x \log^2(x) - 72 x \log^2(x) \nonumber \\
&& -72 x \log(1-x) \log^2(x) - 144 x \log^2(1-x) + 72 \log^2(1-x) + 12 \pi^2 x \log(x) \nonumber \\
&& + 216 x \log(x) + 24 \pi^2 x \log(1-x) \log(x) + 144 x \log(1-x) \log(x) - 72 \log(x+1) \log(x) \nonumber \\
&& -432 x \log(1-x) + 216 \log(1-x) - 3 \pi^2 + 252 \biggr] \nonumber \\
&+& \dfrac{\epsilon^3 x}{90 m_\phi^2 (x-1)^4} \biggl[
12 x \log^5(x) - 45 x \log^4(x) - 90 x \log(1-x) \log^4(x) - 60 x^2 \log^3(x) \nonumber \\
&& + 240 x \log^2(1-x) \log^3(x) + 20 \pi^2 x \log^3(x) + 180 x \log^3(x) \nonumber \\
&& + 120 x \mathrm{HPL}(\{-2\},x) \log^3(x) + 240 x \log(1-x) \log^3(x) - 240 x \log^3(1-x) \log^2(x) \nonumber \\
&& + 540 x^2 \log^2(x) - 360 x \log^2(1-x) \log^2(x) - 45 \pi^2 x \log^2(x) - 720 x \log^2(x) \nonumber \\
&& -360 x \mathrm{HPL}(\{-2\},x) \log^2(x) + 720 x \mathrm{HPL}(\{-2,-1\},x) \log^2(x) \nonumber \\
&& + 360 x^2 \log(1-x) \log^2(x) - 90 \pi^2 x \log(1-x) \log^2(x) - 720 x \log(1-x) \log^2(x) \nonumber \\
&& -720 x \mathrm{HPL}(\{-2\},x) \log(1-x) \log^2(x) - 180 x^2 \log(x+1) \log^2(x) \nonumber \\
&& + 180 \log(x+1) \log^2(x) + 240 x \zeta(3) \log^2(x) - 30 \pi^2 x^2 \log(x) - 2520 x^2 \log(x) \nonumber \\
&& -720 x \mathrm{HPL}(\{-2\},x)^2 \log(x) - 720 x^2 \log^2(1-x) \log(x) \nonumber \\
&& + 120 \pi^2 x \log^2(1-x) \log(x) + 720 x \log^2(1-x) \log(x) \nonumber \\
&& + 1440 x \mathrm{HPL}(\{-2\},x) \log^2(1-x) \log(x) - 360 x^2 \log^2(x+1) \log(x) \nonumber \\
&& + 360 \log^2(x+1) \log(x) + 9 \pi^4 x \log(x) + 90 \pi^2 x \log(x) + 2520 x \log(x) \nonumber \\
&& + 720 x \mathrm{HPL}(\{-4\},x) \log(x) + 360 x^2 \mathrm{HPL}(\{-2\},x) \log(x) \nonumber \\
&& + 60 \pi^2 x \mathrm{HPL}(\{-2\},x) \log(x) + 720 x \mathrm{HPL}(\{-2\},x) \log(x) - 360 \mathrm{HPL}(\{-2\},x) \log(x) \nonumber \\
&& + 1440 x \mathrm{HPL}(\{-3,-1\},x) \log(x) - 1440 x \mathrm{HPL}(\{-2,-1\},x) \log(x) \nonumber \\
&& + 2880 x \mathrm{HPL}(\{-2,-1,-1\},x) \log(x) - 2160 x^2 \log(1-x) \log(x) \nonumber \\
&& + 120 \pi^2 x \log(1-x) \log(x) + 2160 x \log(1-x) \log(x) \nonumber \\
&& + 1440 x \mathrm{HPL}(\{-2\},x) \log(1-x) \log(x) - 2880 x \mathrm{HPL}(\{-2,-1\},x) \log(1-x) \log(x) \nonumber \\
&& + 1080 x^2 \log(x+1) \log(x) + 720 x^2 \log(1-x) \log(x+1) \log(x) \nonumber \\
&& -720 \log(1-x) \log(x+1) \log(x) - 1080 \log(x+1) \log(x) - 360 x \zeta(3) \log(x) \nonumber \\
&& -720 x \log(1-x) \zeta(3) \log(x) + 240 x^2 \log^3(1-x) - 480 x \log^3(1-x) + 240 \log^3(1-x) \nonumber \\
&& + 135 \pi^2 x^2 + 2700 x^2 + 720 x \mathrm{HPL}(\{-2\},x)^2 + 1080 x^2 \log^2(1-x) \nonumber \\
&& -2160 x \log^2(1-x) + 1080 \log^2(1-x) + 5 \pi^4 x - 90 \pi^2 x - 5400 x \nonumber \\
&& -1440 x \mathrm{HPL}(\{-5\},x) - 1080 x^2 \mathrm{HPL}(\{-2\},x) - 120 \pi^2 x \mathrm{HPL}(\{-2\},x) \nonumber \\
&& +1080 \mathrm{HPL}(\{-2\},x) - 8640 x \mathrm{HPL}(\{-4,-1\},x) - 2880 x \mathrm{HPL}(\{-3,-2\},x) \nonumber \\
&& -720 x^2 \mathrm{HPL}(\{-2,-1\},x) + 240 \pi^2 x \mathrm{HPL}(\{-2,-1\},x) \nonumber \\
&& -2880 x \mathrm{HPL}(\{-2\},x) \mathrm{HPL}(\{-2,-1\},x) + 720 \mathrm{HPL}(\{-2,-1\},x) \nonumber \\
&& +11520 x \mathrm{HPL}(\{-3,-1,-1\},x) + 5760 x \mathrm{HPL}(\{-2,-2,-1\},x) \nonumber \\
&& +90 \pi^2 x^2 \log(1-x) + 2520 x^2 \log(1-x) + 1440 x \mathrm{HPL}(\{-2\},x)^2 \log(1-x) \nonumber \\
&& +10 \pi^4 x \log(1-x) - 60 \pi^2 x \log(1-x) - 5040 x \log(1-x) \nonumber \\
&& -720 x^2 \mathrm{HPL}(\{-2\},x) \log(1-x) - 240 \pi^2 x \mathrm{HPL}(\{-2\},x) \log(1-x) \nonumber \\
&& +720 \mathrm{HPL}(\{-2\},x) \log(1-x) - 30 \pi^2 \log(1-x) + 2520 \log(1-x) \nonumber \\
&& -360 \mathrm{HPL}(\{-3\},x) \bigl(x^2 + 8 \log^2(1-x) x + \log^2(x) x - 4 \mathrm{HPL}(\{-2\},x) x \nonumber \\
&& -4 \log(1-x) (\log(x)-2) x - 2 \log(x) x + 4 x - 1\bigr) - 60 \pi^2 x^2 \log(x+1) \nonumber \\
&& +720 x^2 \mathrm{HPL}(\{-2\},x) \log(x+1) - 720 \mathrm{HPL}(\{-2\},x) \log(x+1) + 60 \pi^2 \log(x+1) \nonumber \\
&& +2520 x \zeta(5) + 300 x^2 \zeta(3) + 2160 x \log^2(1-x) \zeta(3) - 30 \pi^2 x \zeta(3) + 1200 x \zeta(3) \nonumber \\
&& -1440 x \mathrm{HPL}(\{-2\},x) \zeta(3) + 2160 x \log(1-x) \zeta(3) - 420 \zeta(3) - 45 \pi^2 + 2700 \biggr] \nonumber \\
&+& {\cal O}(\epsilon^4).
\end{eqnarray}
\endgroup
}
In the above, ${\rm HPL}(\{a_1,a_2,...,a_n\}, x)$ are Harmonic Polylogarithms~\citep{Duhr:2019tlz}. We have shown here terms up to  ${\cal O}(\epsilon^3)$. The ${\cal O}(\epsilon^4)$ term is included in the ancillary files. We note that all the expansion coefficients are finite in the large and small $m_\phi$ limits, as expected. Also, the maximum transcendental weight of HPLs at one-loop is 2 as expected for  $\epsilon^0$ and it increases by 1 with every increase in the order of $\epsilon$.

Unlike the one-loop form factor, the two-loop form factor is both UV and IR divergent. 
After renormalization and IR subtraction, only the finite pieces remain. This finite part of the form factor for the gluon channel defined in Eq.~\ref{eqn:gg2L} is given by,

{\allowdisplaybreaks
\begin{eqnarray}
M_{\mathrm{fin}}^0  &=& \dfrac{x}{90 m_{\phi}^2 N_c (x-1)^5 (x+1)} \Bigg[
2340 N_c^2 x^4 + 540 N_c^2 \log(1-x) x^4 - 540 \log(1-x) x^4 \nonumber \\[4pt]
&&- 540 N_c^2 \log(x) x^4 + 540 \log(x) x^4 - 1260 x^4 - 100 N_c^2 \pi^4 x^3 + 36 \pi^4 x^3 \nonumber \\[4pt]
&&- 45 N_c^2 \log^4(x) x^3 + 15 \log^4(x) x^3 + 1230 N_c^2 \log^3(x) x^3 \nonumber \\[4pt]
&&- 480 N_c^2 \log(1-x) \log^3(x) x^3 - 390 \log^3(x) x^3 - 4680 N_c^2 x^3 - 90 N_c^2 \pi^2 x^3 \nonumber \\[4pt]
&&+ 90 \pi^2 x^3 - 2700 N_c^2 \log^2(x) x^3 - 240 N_c^2 \pi^2 \log^2(x) x^3 + 120 \pi^2 \log^2(x) x^3 \nonumber \\[4pt]
&&+ 1440 N_c^2 \text{HPL}(\{-2\},x) \log^2(x) x^3 - 1440 \text{HPL}(\{-2\},x) \log^2(x) x^3 \nonumber \\[4pt]
&&- 3960 N_c^2 \text{HPL}(\{2\},x) \log^2(x) x^3 + 2520 \text{HPL}(\{2\},x) \log^2(x) x^3 \nonumber \\[4pt]
&&- 4860 N_c^2 \log(1-x) \log^2(x) x^3 + 1980 \log(1-x) \log^2(x) x^3 + 2340 \log^2(x) x^3 \nonumber \\[4pt]
&&+ 1080 N_c^2 \text{HPL}(\{-2\},x) x^3 - 1080 \text{HPL}(\{-2\},x) x^3 + 480 N_c^2 \pi^2 \text{HPL}(\{2\},x) x^3 \nonumber \\[4pt]
&&+ 7920 N_c^2 \text{HPL}(\{3\},x) x^3 - 2160 \text{HPL}(\{3\},x) x^3 - 22320 N_c^2 \text{HPL}(\{4\},x) x^3 \nonumber \\[4pt]
&&+ 19440 \text{HPL}(\{4\},x) x^3 - 11520 N_c^2 \text{HPL}(\{-3,1\},x) x^3 - 5760 N_c^2 \text{HPL}(\{2,-2\},x) x^3 \nonumber \\[4pt]
&&- 11520 N_c^2 \text{HPL}(\{3,-1\},x) x^3 - 1080 N_c^2 \log(1-x) x^3 + 1080 \log(1-x) x^3 \nonumber \\[4pt]
&&+ 1800 N_c^2 \log(x) x^3 + 210 N_c^2 \pi^2 \log(x) x^3 - 210 \pi^2 \log(x) x^3 \nonumber \\[4pt]
&&+ 3960 N_c^2 \text{HPL}(\{-2\},x) \log(x) x^3 - 3960 \text{HPL}(\{-2\},x) \log(x) x^3 \nonumber \\[4pt]
&&- 7200 N_c^2 \text{HPL}(\{2\},x) \log(x) x^3 + 1440 \text{HPL}(\{2\},x) \log(x) x^3 \nonumber \\[4pt]
&&+ 14400 N_c^2 \text{HPL}(\{3\},x) \log(x) x^3 - 11520 \text{HPL}(\{3\},x) \log(x) x^3 \nonumber \\[4pt]
&&+ 5760 N_c^2 \text{HPL}(\{-2,1\},x) \log(x) x^3 + 5760 N_c^2 \text{HPL}(\{2,-1\},x) \log(x) x^3 \nonumber \\[4pt]
&&+ 2160 N_c^2 \log(1-x) \log(x) x^3 + 480 N_c^2 \pi^2 \log(1-x) \log(x) x^3 \nonumber \\[4pt]
&&+ 5760 N_c^2 \text{HPL}(\{-2\},x) \log(1-x) \log(x) x^3 - 2160 \log(1-x) \log(x) x^3 \nonumber \\[4pt]
&&- 1800 \log(x) x^3 - 1080 N_c^2 \log(x) \log(x+1) x^3 + 1080 \log(x) \log(x+1) x^3 \nonumber \\[4pt]
&&- 1980 N_c^2 \zeta(3) x^3 + 8640 N_c^2 \log(1-x) \zeta(3) x^3 - 7200 N_c^2 \log(x) \zeta(3) x^3 \nonumber \\[4pt]
&&+ 2880 \log(x) \zeta(3) x^3 - 3780 \zeta(3) x^3 + 2520 x^3 + 540 N_c^2 \log^3(x) x^2 \nonumber \\[4pt]
&&+ 180 \log^3(x) x^2 + 180 N_c^2 \pi^2 x^2 - 180 \pi^2 x^2 + 1620 N_c^2 \log^2(x) x^2 \nonumber \\[4pt]
&&- 1620 \log^2(x) x^2 - 2160 N_c^2 \text{HPL}(\{-2\},x) x^2 + 2160 \text{HPL}(\{-2\},x) x^2 \nonumber \\[4pt]
&&- 1980 N_c^2 \log(x) x^2 - 4320 N_c^2 \log(1-x) \log(x) x^2 + 4320 \log(1-x) \log(x) x^2 \nonumber \\[4pt]
&&+ 1980 \log(x) x^2 + 2160 N_c^2 \log(x) \log(x+1) x^2 - 2160 \log(x) \log(x+1) x^2 \nonumber \\[4pt]
&&+ 28 N_c^2 \pi^4 x + 36 \pi^4 x + 15 N_c^2 \log^4(x) x + 15 \log^4(x) x \nonumber \\[4pt]
&&- 690 N_c^2 \log^3(x) x + 480 N_c^2 \log(1-x) \log^3(x) x + 570 \log^3(x) x \nonumber \\[4pt]
&&+ 4680 N_c^2 x - 90 N_c^2 \pi^2 x + 90 \pi^2 x + 1080 N_c^2 \log^2(x) x \nonumber \\[4pt]
&&+ 120 \pi^2 \log^2(x) x + 1440 N_c^2 \text{HPL}(\{-2\},x) \log^2(x) x - 1440 \text{HPL}(\{-2\},x) \log^2(x) x \nonumber \\[4pt]
&&- 1080 N_c^2 \text{HPL}(\{2\},x) \log^2(x) x + 2520 \text{HPL}(\{2\},x) \log^2(x) x \nonumber \\[4pt]
&&+ 4860 N_c^2 \log(1-x) \log^2(x) x - 1980 \log(1-x) \log^2(x) x - 720 \log^2(x) x \nonumber \\[4pt]
&&+ 8640 \left(N_c^2 (4 x^2 + 2) - 3 (x^2 + 1)\right) \text{HPL}(\{-4\},x) x \nonumber \\[4pt]
&&+ 1080 N_c^2 \text{HPL}(\{-2\},x) x - 1080 \text{HPL}(\{-2\},x) x - 480 N_c^2 \pi^2 \text{HPL}(\{2\},x) x \nonumber \\[4pt]
&&- 7920 N_c^2 \text{HPL}(\{3\},x) x + 2160 \text{HPL}(\{3\},x) x - 16560 N_c^2 \text{HPL}(\{4\},x) x \nonumber \\[4pt]
&&+ 19440 \text{HPL}(\{4\},x) x + 11520 N_c^2 \text{HPL}(\{-3,1\},x) x + 5760 N_c^2 \text{HPL}(\{2,-2\},x) x \nonumber \\[4pt]
&&+ 11520 N_c^2 \text{HPL}(\{3,-1\},x) x + 1080 N_c^2 \log(1-x) x - 1080 \log(1-x) x \nonumber \\[4pt]
&&+ 720 N_c^2 \log(x) x - 210 N_c^2 \pi^2 \log(x) x + 210 \pi^2 \log(x) x \nonumber \\[4pt]
&&- 3960 N_c^2 \text{HPL}(\{-2\},x) \log(x) x + 3960 \text{HPL}(\{-2\},x) \log(x) x \nonumber \\[4pt]
&&+ 7200 N_c^2 \text{HPL}(\{2\},x) \log(x) x - 1440 \text{HPL}(\{2\},x) \log(x) x \nonumber \\[4pt]
&&+ 8640 N_c^2 \text{HPL}(\{3\},x) \log(x) x - 11520 \text{HPL}(\{3\},x) \log(x) x \nonumber \\[4pt]
&&- 5760 N_c^2 \text{HPL}(\{-2,1\},x) \log(x) x - 5760 N_c^2 \text{HPL}(\{2,-1\},x) \log(x) x \nonumber \\[4pt]
&&+ 2160 N_c^2 \log(1-x) \log(x) x - 480 N_c^2 \pi^2 \log(1-x) \log(x) x \nonumber \\[4pt]
&&- 5760 N_c^2 \text{HPL}(\{-2\},x) \log(1-x) \log(x) x - 2160 \log(1-x) \log(x) x \nonumber \\[4pt]
&&- 720 \log(x) x - 720 \text{HPL}(\{-3\},x) \Big(16 (x^2 - 1) \log(1-x) N_c^2 \nonumber \\[4pt]
&&\quad + 11 (N_c^2 - 1) (x^2 - 1) + 4 \left(N_c^2 (5 x^2 + 3) - 4 (x^2 + 1)\right) \log(x)\Big) x \nonumber \\[4pt]
&&- 1080 N_c^2 \log(x) \log(x+1) x + 1080 \log(x) \log(x+1) x \nonumber \\[4pt]
&&+ 1980 N_c^2 \zeta(3) x - 8640 N_c^2 \log(1-x) \zeta(3) x + 1440 N_c^2 \log(x) \zeta(3) x \nonumber \\[4pt]
&&+ 2880 \log(x) \zeta(3) x + 3780 \zeta(3) x - 2520 x - 2340 N_c^2 - 540 N_c^2 \log(1-x) \nonumber \\[4pt]
&&+ 540 \log(1-x) + 1260 \Bigg] \label{eq:Fgg}
\end{eqnarray}
}

As we can see the two-loop form factor contains HPLs up to weight 4. 
Note that unlike the one-loop form factor, the two-loop form factor has non-trivial color factor dependence. The two-loop form factor has color structure given by $\Big( N_c {\cal A}(x) + \dfrac{1}{N_c} {\cal B}(x)\Big)$. The subleading color piece ${\cal B}(x)$ arise from non-planar two-loop Feynman integrals. We have evaluated two-loop form factor at $\mathcal{O}(\epsilon)$. and it is included in the ancillary files.

\subsection{Quark channel}
As discussed earlier in section~\ref{sec:qqchannel}, the quark-channel contribution is two-loop at leading order. 
The form factor in terms of massless ratios $w$ and $z$ is given by, 
{\allowdisplaybreaks
\small
\begin{eqnarray}
\mathcal{F}_{qq}^{\rm 2L} &=& i~C_F~m_q~\dfrac{16}{m_\phi^2}~\lambda_d~ \Big(\dfrac{\alpha_s}{4 \pi}\Big)^2 \nonumber \\
&&\Bigg[
\dfrac{\pi^2 (z^2-1)^3 G(-1,0,z) w^4}{3 (w^2-1)^4 (z^2+1)^3}
+ \dfrac{2 (3 w^2+1) (z^2-1)^2 G(0,0,w) w^4}{(w^2-1)^4 (z^2+1)^2}
+ \dfrac{10 \pi^2 z^2 (z^2-1) G\left(0,-\dfrac{1}{z},w\right) w^4}{3 (w^2-1)^4 (z^2+1)^3} \nonumber \\
&& - \dfrac{2 \pi^2 z^2 (z^2-1) G\left(0,\dfrac{1}{z},w\right) w^4}{3 (w^2-1)^4 (z^2+1)^3}
+ \dfrac{10 \pi^2 z^2 (z^2-1) G(0,-z,w) w^4}{3 (w^2-1)^4 (z^2+1)^3}
- \dfrac{2 \pi^2 z^2 (z^2-1) G(0,z,w) w^4}{3 (w^2-1)^4 (z^2+1)^3} \nonumber \\
&& + \dfrac{\pi^2 (z^2-1)^3 G(1,0,z) w^4}{3 (w^2-1)^4 (z^2+1)^3}
+ \dfrac{8 (z^2-1)^2 G(-1,0,0,w) w^4}{(w^2-1)^4 (z^2+1)^2}
- \dfrac{4 (z^2-1)^2 G(0,-1,0,w) w^4}{(w^2-1)^4 (z^2+1)^2} \nonumber \\
&& + \dfrac{4 (z^2-1)^2 G(0,0,-1,z) w^4}{(w^2-1)^4 (z^2+1)^2}
- \dfrac{4 (z^2-1)^2 G(0,0,0,z) w^4}{(w^2-1)^4 (z^2+1)^2}
+ \dfrac{4 (z^2-1)^2 G(0,0,1,z) w^4}{(w^2-1)^4 (z^2+1)^2} \nonumber \\
&& - \dfrac{8 (z^2-1)^2 G(0,-i,0,w) w^4}{(w^2-1)^4 (z^2+1)^2}
- \dfrac{8 (z^2-1)^2 G(0,i,0,w) w^4}{(w^2-1)^4 (z^2+1)^2}
- \dfrac{4 (z^2-1)^2 G(0,1,0,w) w^4}{(w^2-1)^4 (z^2+1)^2} \nonumber \\
&& + \dfrac{2 (z^2-1) (z^4-2 w^2 z^2+1) G\left(0,-\dfrac{1}{z},-1,w\right) w^4}{(w^2-1)^4 (z^2+1)^3}
- \dfrac{2 (z^2-1) (z^4-2 w^2 z^2+1) G\left(0,-\dfrac{1}{z},0,w\right) w^4}{(w^2-1)^4 (z^2+1)^3} \nonumber \\
&& + \dfrac{2 (z^2-1) (z^4-2 w^2 z^2+1) G\left(0,-\dfrac{1}{z},1,w\right) w^4}{(w^2-1)^4 (z^2+1)^3}
+ \dfrac{2 (z^2-1) (z^4-2 w^2 z^2+1) G\left(0,\dfrac{1}{z},-1,w\right) w^4}{(w^2-1)^4 (z^2+1)^3} \nonumber \\
&& - \dfrac{2 (z^2-1) (z^4-2 w^2 z^2+1) G\left(0,\dfrac{1}{z},0,w\right) w^4}{(w^2-1)^4 (z^2+1)^3}
+ \dfrac{2 (z^2-1) (z^4-2 w^2 z^2+1) G\left(0,\dfrac{1}{z},1,w\right) w^4}{(w^2-1)^4 (z^2+1)^3} \nonumber \\
&& - \dfrac{2 (z^2-1) (z^4-2 w^2 z^2+1) G(0,-z,-1,w) w^4}{(w^2-1)^4 (z^2+1)^3}
+ \dfrac{2 (z^2-1) (z^4-2 w^2 z^2+1) G(0,-z,0,w) w^4}{(w^2-1)^4 (z^2+1)^3} \nonumber \\
&& - \dfrac{2 (z^2-1) (z^4-2 w^2 z^2+1) G(0,-z,1,w) w^4}{(w^2-1)^4 (z^2+1)^3}
- \dfrac{2 (z^2-1) (z^4-2 w^2 z^2+1) G(0,z,-1,w) w^4}{(w^2-1)^4 (z^2+1)^3} \nonumber \\
&& + \dfrac{2 (z^2-1) (z^4-2 w^2 z^2+1) G(0,z,0,w) w^4}{(w^2-1)^4 (z^2+1)^3}
- \dfrac{2 (z^2-1) (z^4-2 w^2 z^2+1) G(0,z,1,w) w^4}{(w^2-1)^4 (z^2+1)^3} \nonumber \\
&& + \dfrac{8 (z^2-1)^2 G(1,0,0,w) w^4}{(w^2-1)^4 (z^2+1)^2}
+ \dfrac{4 (z^2-1)^3 G(-1,0,-1,-1,z) w^4}{(w^2-1)^4 (z^2+1)^3} \nonumber \\
&& - \dfrac{4 (z^2-1)^3 (G(-1,0,-1,0,z)+G(-1,0,-1,1,z)-G(-1,0,0,-1,z)) w^4}{(w^2-1)^4 (z^2+1)^3}
 \nonumber \\
&& + \dfrac{4 (z^2-1)^3 (G(-1,0,0,0,z)-G(-1,0,0,1,z)+G(-1,0,1,-1,z)) w^4}{(w^2-1)^4 (z^2+1)^3}
\nonumber \\
&& - \dfrac{4 (z^2-1)^3 G(-1,0,1,0,z) w^4}{(w^2-1)^4 (z^2+1)^3}
+ \dfrac{4 (z^2-1)^3 G(-1,0,1,1,z) w^4}{(w^2-1)^4 (z^2+1)^3} \nonumber \\
&& + \dfrac{2 (z^2-1)^3 (G\left(-1,0,-\dfrac{1}{z},-1,w\right)-G\left(-1,0,-\dfrac{1}{z},0,w\right)+G\left(-1,0,-\dfrac{1}{z},1,w\right)) w^4}{(w^2-1)^4 (z^2+1)^3}\nonumber \\
&& + \dfrac{2 (z^2-1)^3 (G\left(-1,0,\dfrac{1}{z},-1,w\right)-G\left(-1,0,\dfrac{1}{z},0,w\right)+G\left(-1,0,\dfrac{1}{z},1,w\right)) w^4}{(w^2-1)^4 (z^2+1)^3} \nonumber \\
&& - \dfrac{2 (z^2-1)^3 (G(-1,0,-z,-1,w)+ G(-1,0,-z,0,w)-G(-1,0,-z,1,w)) w^4}{(w^2-1)^4 (z^2+1)^3}\nonumber \\
&& - \dfrac{2 (z^2-1)^3 (G(-1,0,z,-1,w)+G(-1,0,z,0,w)-G(-1,0,z,1,w)) w^4}{(w^2-1)^4 (z^2+1)^3} \nonumber \\
&& + \dfrac{2 (z^2-1)^3 G\left(-1,-\dfrac{1}{z},-1,0,w\right) w^4}{(w^2-1)^4 (z^2+1)^3}
+ \dfrac{2 (z^2-1)^3 G\left(-1,-\dfrac{1}{z},0,-1,w\right) w^4}{(w^2-1)^4 (z^2+1)^3} \nonumber \\
&& - \dfrac{4 (z^2-1)^3 G\left(-1,-\dfrac{1}{z},0,0,w\right) w^4}{(w^2-1)^4 (z^2+1)^3}
+ \dfrac{2 (z^2-1)^3 G\left(-1,-\dfrac{1}{z},0,1,w\right) w^4}{(w^2-1)^4 (z^2+1)^3} \nonumber \\
&& + \dfrac{2 (z^2-1)^3 G\left(-1,-\dfrac{1}{z},1,0,w\right) w^4}{(w^2-1)^4 (z^2+1)^3}
+ \dfrac{2 (z^2-1)^3 G\left(-1,\dfrac{1}{z},-1,0,w\right) w^4}{(w^2-1)^4 (z^2+1)^3} \nonumber \\
&& + \dfrac{2 (z^2-1)^3 G\left(-1,\dfrac{1}{z},0,-1,w\right) w^4}{(w^2-1)^4 (z^2+1)^3}
- \dfrac{4 (z^2-1)^3 G\left(-1,\dfrac{1}{z},0,0,w\right) w^4}{(w^2-1)^4 (z^2+1)^3} \nonumber \\
&& + \dfrac{2 (z^2-1)^3 G\left(-1,\dfrac{1}{z},0,1,w\right) w^4}{(w^2-1)^4 (z^2+1)^3}
+ \dfrac{2 (z^2-1)^3 G\left(-1,\dfrac{1}{z},1,0,w\right) w^4}{(w^2-1)^4 (z^2+1)^3} \nonumber \\
&& - \dfrac{2 (z^2-1)^3 G(-1,-z,-1,0,w) w^4}{(w^2-1)^4 (z^2+1)^3}
- \dfrac{2 (z^2-1)^3 G(-1,-z,0,-1,w) w^4}{(w^2-1)^4 (z^2+1)^3} \nonumber \\
&& + \dfrac{4 (z^2-1)^3 G(-1,-z,0,0,w) w^4}{(w^2-1)^4 (z^2+1)^3}
- \dfrac{2 (z^2-1)^3 G(-1,-z,0,1,w) w^4}{(w^2-1)^4 (z^2+1)^3} \nonumber \\
&& - \dfrac{2 (z^2-1)^3 G(-1,-z,1,0,w) w^4}{(w^2-1)^4 (z^2+1)^3}
- \dfrac{2 (z^2-1)^3 G(-1,z,-1,0,w) w^4}{(w^2-1)^4 (z^2+1)^3} \nonumber \\
&& - \dfrac{2 (z^2-1)^3 G(-1,z,0,-1,w) w^4}{(w^2-1)^4 (z^2+1)^3}
+ \dfrac{4 (z^2-1)^3 G(-1,z,0,0,w) w^4}{(w^2-1)^4 (z^2+1)^3} \nonumber \\
&& - \dfrac{2 (z^2-1)^3 G(-1,z,0,1,w) w^4}{(w^2-1)^4 (z^2+1)^3}
- \dfrac{2 (z^2-1)^3 G(-1,z,1,0,w) w^4}{(w^2-1)^4 (z^2+1)^3} \nonumber \\
&& - \dfrac{4 (z^2-1)^3 G(0,0,-1,-1,z) w^4}{(w^2-1)^4 (z^2+1)^3}
+ \dfrac{4 (z^2-1)^3 G(0,0,-1,0,z) w^4}{(w^2-1)^4 (z^2+1)^3} \nonumber \\
&& - \dfrac{4 (z^2-1)^3 G(0,0,-1,1,z) w^4}{(w^2-1)^4 (z^2+1)^3}
+ \dfrac{4 (z^2-1) G(0,0,0,-1,z) w^4}{(w^2-1)^4 (z^2+1)} \nonumber \\
&& - \dfrac{4 (z^2-1) G(0,0,0,0,z) w^4}{(w^2-1)^4 (z^2+1)}
+ \dfrac{4 (z^2-1) G(0,0,0,1,z) w^4}{(w^2-1)^4 (z^2+1)} \nonumber \\
&& - \dfrac{4 (z^2-1)^3 G(0,0,1,-1,z) w^4}{(w^2-1)^4 (z^2+1)^3}
+ \dfrac{4 (z^2-1)^3 G(0,0,1,0,z) w^4}{(w^2-1)^4 (z^2+1)^3} \nonumber \\
&& - \dfrac{4 (z^2-1)^3 G(0,0,1,1,z) w^4}{(w^2-1)^4 (z^2+1)^3}+ \dfrac{2 (z^2-1) G\left(0,0,-\dfrac{1}{z},0,w\right) w^4}{(w^2-1)^4 (z^2+1)} - \dfrac{2 (z^2-1) G\left(0,0,-\dfrac{1}{z},-1,w\right) w^4}{(w^2-1)^4 (z^2+1)} \nonumber \\
&&- \dfrac{2 (z^2-1) G\left(0,0,-\dfrac{1}{z},1,w\right) w^4}{(w^2-1)^4 (z^2+1)}
- \dfrac{2 (z^2-1) G\left(0,0,\dfrac{1}{z},-1,w\right) w^4}{(w^2-1)^4 (z^2+1)}
+ \dfrac{2 (z^2-1) G\left(0,0,\dfrac{1}{z},0,w\right) w^4}{(w^2-1)^4 (z^2+1)} \nonumber \\
&& - \dfrac{2 (z^2-1) G\left(0,0,\dfrac{1}{z},1,w\right) w^4}{(w^2-1)^4 (z^2+1)}
+ \dfrac{2 (z^2-1) G(0,0,-z,-1,w) w^4}{(w^2-1)^4 (z^2+1)}
- \dfrac{2 (z^2-1) G(0,0,-z,0,w) w^4}{(w^2-1)^4 (z^2+1)} \nonumber \\
&& + \dfrac{2 (z^2-1) (G(0,0,-z,1,w)+G(0,0,z,-1,w)-G(0,0,z,0,w) ) w^4}{(w^2-1)^4 (z^2+1)}
\nonumber \\
&& + \dfrac{2 (z^2-1) G(0,0,z,1,w) w^4}{(w^2-1)^4 (z^2+1)}
+ \dfrac{16 z^2 (z^2-1) G\left(0,-i,-\dfrac{1}{z},-1,w\right) w^4}{(w^2-1)^4 (z^2+1)^3} \nonumber \\
&& - \dfrac{16 z^2 (z^2-1) G\left(0,-i,-\dfrac{1}{z},0,w\right) w^4}{(w^2-1)^4 (z^2+1)^3}
+ \dfrac{16 z^2 (z^2-1) G\left(0,-i,-\dfrac{1}{z},1,w\right) w^4}{(w^2-1)^4 (z^2+1)^3} \nonumber \\
&& + \dfrac{16 z^2 (z^2-1) G\left(0,-i,\dfrac{1}{z},-1,w\right) w^4}{(w^2-1)^4 (z^2+1)^3}
- \dfrac{16 z^2 (z^2-1) G\left(0,-i,\dfrac{1}{z},0,w\right) w^4}{(w^2-1)^4 (z^2+1)^3} \nonumber \\
&& + \dfrac{16 z^2 (z^2-1) G\left(0,-i,\dfrac{1}{z},1,w\right) w^4}{(w^2-1)^4 (z^2+1)^3}
- \dfrac{16 z^2 (z^2-1) G(0,-i,-z,-1,w) w^4}{(w^2-1)^4 (z^2+1)^3} \nonumber \\
&& + \dfrac{16 z^2 (z^2-1) G(0,-i,-z,0,w) w^4}{(w^2-1)^4 (z^2+1)^3}
- \dfrac{16 z^2 (z^2-1) G(0,-i,-z,1,w) w^4}{(w^2-1)^4 (z^2+1)^3} \nonumber \\
&& - \dfrac{16 z^2 (z^2-1) G(0,-i,z,-1,w) w^4}{(w^2-1)^4 (z^2+1)^3}
+ \dfrac{16 z^2 (z^2-1) G(0,-i,z,0,w) w^4}{(w^2-1)^4 (z^2+1)^3} \nonumber \\
&& - \dfrac{16 z^2 (z^2-1) G(0,-i,z,1,w) w^4}{(w^2-1)^4 (z^2+1)^3}
+ \dfrac{16 z^2 (z^2-1) G\left(0,i,-\dfrac{1}{z},-1,w\right) w^4}{(w^2-1)^4 (z^2+1)^3} \nonumber \\
&& - \dfrac{16 z^2 (z^2-1) G\left(0,i,-\dfrac{1}{z},0,w\right) w^4}{(w^2-1)^4 (z^2+1)^3}
+ \dfrac{16 z^2 (z^2-1) G\left(0,i,-\dfrac{1}{z},1,w\right) w^4}{(w^2-1)^4 (z^2+1)^3} \nonumber \\
&& + \dfrac{16 z^2 (z^2-1) G\left(0,i,\dfrac{1}{z},-1,w\right) w^4}{(w^2-1)^4 (z^2+1)^3}
- \dfrac{16 z^2 (z^2-1) G\left(0,i,\dfrac{1}{z},0,w\right) w^4}{(w^2-1)^4 (z^2+1)^3} \nonumber \\
&& + \dfrac{16 z^2 (z^2-1) G\left(0,i,\dfrac{1}{z},1,w\right) w^4}{(w^2-1)^4 (z^2+1)^3}
- \dfrac{16 z^2 (z^2-1) G(0,i,-z,-1,w) w^4}{(w^2-1)^4 (z^2+1)^3} \nonumber \\
&& + \dfrac{16 z^2 (z^2-1) G(0,i,-z,0,w) w^4}{(w^2-1)^4 (z^2+1)^3}
- \dfrac{16 z^2 (z^2-1) G(0,i,-z,1,w) w^4}{(w^2-1)^4 (z^2+1)^3} \nonumber \\
&& - \dfrac{16 z^2 (z^2-1) G(0,i,z,-1,w) w^4}{(w^2-1)^4 (z^2+1)^3}
+ \dfrac{16 z^2 (z^2-1) G(0,i,z,0,w) w^4}{(w^2-1)^4 (z^2+1)^3} \nonumber \\
&&- \dfrac{2 (z^2-1)^3 G\left(0,-\dfrac{1}{z},-1,0,w\right) w^4}{(w^2-1)^4 (z^2+1)^3}
- \dfrac{2 (z^2-1)^3 G\left(0,-\dfrac{1}{z},0,-1,w\right) w^4}{(w^2-1)^4 (z^2+1)^3} \nonumber \\
&& + \dfrac{4 (z^2-1) (z^4+1) G\left(0,-\dfrac{1}{z},0,0,w\right) w^4}{(w^2-1)^4 (z^2+1)^3}
- \dfrac{2 (z^2-1)^3 G\left(0,-\dfrac{1}{z},0,1,w\right) w^4}{(w^2-1)^4 (z^2+1)^3} \nonumber \\
&& - \dfrac{2 (z^2-1)^3 G\left(0,-\dfrac{1}{z},1,0,w\right) w^4}{(w^2-1)^4 (z^2+1)^3}
- \dfrac{8 z^2 (z^2-1) G\left(0,-\dfrac{1}{z},\dfrac{1}{z},-1,w\right) w^4}{(w^2-1)^4 (z^2+1)^3} \nonumber \\
&& + \dfrac{8 z^2 (z^2-1) G\left(0,-\dfrac{1}{z},\dfrac{1}{z},0,w\right) w^4}{(w^2-1)^4 (z^2+1)^3}
- \dfrac{8 z^2 (z^2-1) G\left(0,-\dfrac{1}{z},\dfrac{1}{z},1,w\right) w^4}{(w^2-1)^4 (z^2+1)^3} \nonumber \\
&& + \dfrac{8 z^2 (z^2-1) G\left(0,-\dfrac{1}{z},z,-1,w\right) w^4}{(w^2-1)^4 (z^2+1)^3}
- \dfrac{8 z^2 (z^2-1) G\left(0,-\dfrac{1}{z},z,0,w\right) w^4}{(w^2-1)^4 (z^2+1)^3} \nonumber \\
&& + \dfrac{8 z^2 (z^2-1) G\left(0,-\dfrac{1}{z},z,1,w\right) w^4}{(w^2-1)^4 (z^2+1)^3}
- \dfrac{2 (z^2-1)^3 G\left(0,\dfrac{1}{z},-1,0,w\right) w^4}{(w^2-1)^4 (z^2+1)^3} \nonumber \\
&& - \dfrac{2 (z^2-1)^3 G\left(0,\dfrac{1}{z},0,-1,w\right) w^4}{(w^2-1)^4 (z^2+1)^3}
+ \dfrac{4 (z^2-1) (z^4+1) G\left(0,\dfrac{1}{z},0,0,w\right) w^4}{(w^2-1)^4 (z^2+1)^3} \nonumber \\
&& - \dfrac{2 (z^2-1)^3 G\left(0,\dfrac{1}{z},0,1,w\right) w^4}{(w^2-1)^4 (z^2+1)^3}
- \dfrac{2 (z^2-1)^3 G\left(0,\dfrac{1}{z},1,0,w\right) w^4}{(w^2-1)^4 (z^2+1)^3} \nonumber \\
&& - \dfrac{8 z^2 (z^2-1) G\left(0,\dfrac{1}{z},-\dfrac{1}{z},-1,w\right) w^4}{(w^2-1)^4 (z^2+1)^3}
+ \dfrac{8 z^2 (z^2-1) G\left(0,\dfrac{1}{z},-\dfrac{1}{z},0,w\right) w^4}{(w^2-1)^4 (z^2+1)^3} \nonumber \\
&& - \dfrac{8 z^2 (z^2-1) G\left(0,\dfrac{1}{z},-\dfrac{1}{z},1,w\right) w^4}{(w^2-1)^4 (z^2+1)^3}
+ \dfrac{8 z^2 (z^2-1) G\left(0,\dfrac{1}{z},-z,-1,w\right) w^4}{(w^2-1)^4 (z^2+1)^3} \nonumber \\
&& - \dfrac{8 z^2 (z^2-1) G\left(0,\dfrac{1}{z},-z,0,w\right) w^4}{(w^2-1)^4 (z^2+1)^3}
+ \dfrac{8 z^2 (z^2-1) G\left(0,\dfrac{1}{z},-z,1,w\right) w^4}{(w^2-1)^4 (z^2+1)^3} \nonumber \\
&& + \dfrac{2 (z^2-1)^3 G(0,-z,-1,0,w) w^4}{(w^2-1)^4 (z^2+1)^3}
+ \dfrac{2 (z^2-1)^3 G(0,-z,0,-1,w) w^4}{(w^2-1)^4 (z^2+1)^3} \nonumber \\
&& - \dfrac{4 (z^2-1) (z^4+1) G(0,-z,0,0,w) w^4}{(w^2-1)^4 (z^2+1)^3}
+ \dfrac{2 (z^2-1)^3 G(0,-z,0,1,w) w^4}{(w^2-1)^4 (z^2+1)^3} \nonumber \\
&& + \dfrac{2 (z^2-1)^3 G(0,-z,1,0,w) w^4}{(w^2-1)^4 (z^2+1)^3}
- \dfrac{8 z^2 (z^2-1) G\left(0,-z,\dfrac{1}{z},-1,w\right) w^4}{(w^2-1)^4 (z^2+1)^3} \nonumber \\
&& + \dfrac{8 z^2 (z^2-1) G\left(0,-z,\dfrac{1}{z},0,w\right) w^4}{(w^2-1)^4 (z^2+1)^3}
- \dfrac{8 z^2 (z^2-1) G\left(0,-z,\dfrac{1}{z},1,w\right) w^4}{(w^2-1)^4 (z^2+1)^3} \nonumber \\
&& + \dfrac{8 z^2 (z^2-1) G(0,-z,z,-1,w) w^4}{(w^2-1)^4 (z^2+1)^3}
- \dfrac{8 z^2 (z^2-1) G(0,-z,z,0,w) w^4}{(w^2-1)^4 (z^2+1)^3} \nonumber \\
&& + \dfrac{8 z^2 (z^2-1) G(0,-z,z,1,w) w^4}{(w^2-1)^4 (z^2+1)^3}
+ \dfrac{2 (z^2-1)^3 G(0,z,-1,0,w) w^4}{(w^2-1)^4 (z^2+1)^3} \nonumber \\
&& + \dfrac{2 (z^2-1)^3 G(0,z,0,-1,w) w^4}{(w^2-1)^4 (z^2+1)^3}- \dfrac{16 z^2 (z^2-1) G(0,i,z,1,w) w^4}{(w^2-1)^4 (z^2+1)^3} \nonumber \\
&&- \dfrac{8 z^2 (z^2-1) G(0,-z,z,w) w^4}{(w^2-1)^4 (z^2+1)^3}
- \dfrac{2 (z^2-1)^3 G(0,z,0,w) w^4}{(w^2-1)^4 (z^2+1)^3} \nonumber \\
&& + \dfrac{8 z^2 (z^2-1) G\left(0,z,-\dfrac{1}{z},w\right) w^4}{(w^2-1)^4 (z^2+1)^3}
- \dfrac{8 z^2 (z^2-1) G(0,z,-z,w) w^4}{(w^2-1)^4 (z^2+1)^3} \nonumber \\
&& - \dfrac{2 (z^2-1)^3 G\left(1,0,-\dfrac{1}{z},w\right) w^4}{(w^2-1)^4 (z^2+1)^3}
- \dfrac{2 (z^2-1)^3 G\left(1,0,\dfrac{1}{z},w\right) w^4}{(w^2-1)^4 (z^2+1)^3} \nonumber \\
&& + \dfrac{2 (z^2-1)^3 G(1,0,-z,w) w^4}{(w^2-1)^4 (z^2+1)^3}
+ \dfrac{2 (z^2-1)^3 G(1,0,z,w) w^4}{(w^2-1)^4 (z^2+1)^3} \nonumber \\
&& - \dfrac{2 (z^2-1)^3 G\left(1,-\dfrac{1}{z},0,w\right) w^4}{(w^2-1)^4 (z^2+1)^3}
- \dfrac{2 (z^2-1)^3 G\left(1,\dfrac{1}{z},0,w\right) w^4}{(w^2-1)^4 (z^2+1)^3} \nonumber \\
&& + \dfrac{2 (z^2-1)^3 G(1,-z,0,w) w^4}{(w^2-1)^4 (z^2+1)^3}
+ \dfrac{2 (z^2-1)^3 G(1,z,0,w) w^4}{(w^2-1)^4 (z^2+1)^3} \nonumber \\
&& + \dfrac{2 (w-z) z (w z-1) (z^2-1) G\left(\dfrac{1}{z},w\right) w^2}{(w^2-1)^3 (z^2+1)^3}
- \dfrac{2 z (w+z) (w z+1) (z^2-1) G(-z,w) w^2}{(w^2-1)^3 (z^2+1)^3} \nonumber \\
&& - \dfrac{2 (w-z) z (w z-1) (z^2-1) G(z,w) w^2}{(w^2-1)^3 (z^2+1)^3}
- \dfrac{4 (w^2+1) z^2 (z^2-1) G\left(-i,-\dfrac{1}{z},w\right) w^2}{(w^2-1)^3 (z^2+1)^3} \nonumber \\
&& - \dfrac{4 (w^2+1) z^2 (z^2-1) G\left(-i,\dfrac{1}{z},w\right) w^2}{(w^2-1)^3 (z^2+1)^3}
+ \dfrac{4 (w^2+1) z^2 (z^2-1) G(-i,-z,w) w^2}{(w^2-1)^3 (z^2+1)^3} \nonumber \\
&& + \dfrac{4 (w^2+1) z^2 (z^2-1) G(-i,z,w) w^2}{(w^2-1)^3 (z^2+1)^3}
- \dfrac{4 (w^2+1) z^2 (z^2-1) G\left(i,-\dfrac{1}{z},w\right) w^2}{(w^2-1)^3 (z^2+1)^3} \nonumber \\
&& - \dfrac{4 (w^2+1) z^2 (z^2-1) (G\left(i,\dfrac{1}{z},w\right)+G(i,-z,w)+G(i,z,w)) w^2}{(w^2-1)^3 (z^2+1)^3}\nonumber \\
&& + \dfrac{2 (w-z) (w+z) (w z-1) (w z+1) (z^2-1) G\left(-\dfrac{1}{z},0,w\right) w^2}{(w^2-1)^4 (z^2+1)^3} \nonumber \\
&& + \dfrac{2 (w z-1) (w z+1) (z^2-1) (w^2+z^2) (G\left(-\dfrac{1}{z},\dfrac{1}{z},w\right)- G\left(-\dfrac{1}{z},z,w\right)) w^2}{(w^2-1)^4 (z^2+1)^3} \nonumber \\
&& + \dfrac{2 (w-z) (w+z) (w z-1) (w z+1) (z^2-1)( G\left(\dfrac{1}{z},0,w\right)+G\left(\dfrac{1}{z},-\dfrac{1}{z},w\right)) w^2}{(w^2-1)^4 (z^2+1)^3} \nonumber \\
&& - \dfrac{2 (w z-1) (w z+1) (z^2-1) (w^2+z^2) G\left(\dfrac{1}{z},-z,w\right) w^2}{(w^2-1)^4 (z^2+1)^3} \nonumber \\
&& - \dfrac{2 (w-z) (w+z) (w z-1) (w z+1) (z^2-1) G(-z,0,w) w^2}{(w^2-1)^4 (z^2+1)^3} \nonumber \\
&& + \dfrac{2 (w-z) (w+z) (z^2-1) (w^2 z^2+1) (G\left(-z,\dfrac{1}{z},w\right) +G(-z,z,w)) w^2}{(w^2-1)^4 (z^2+1)^3} \nonumber \\
&& - \dfrac{2 (w-z) (w+z) (w z-1) (w z+1) (z^2-1) G(z,0,w) w^2}{(w^2-1)^4 (z^2+1)^3} \nonumber \\
&& + \dfrac{2 (w-z) (w+z) (z^2-1) (w^2 z^2+1) (G\left(z,-\dfrac{1}{z},w\right)-G(z,-z,w)) w^2}{(w^2-1)^4 (z^2+1)^3} \nonumber \\
&&+ \dfrac{(z^2-1)^2 \left((w^2-1)^2 (z^2+1)-w^2 (z^2-1) \zeta(3)\right) w^2}{(w^2-1)^4 (z^2+1)^3}
\Bigg] \label{eq:Fqq}
\end{eqnarray}
}
The amplitude is directly proportional to $m_q$, the mass of the heavy quark. It has been expressed in terms of Generalized Polylogarithms (GPL) $G(a_1,a_2,...,a_n,z)$~\citep{Duhr:2019tlz}, which are functions of two massless ratios $w$ and $z$ and are of maximum weight or transcendentality 4 as expected at two-loop order.

\section{Conclusions and outlook} 
\label{sec:conclusions}
In this work, we have computed the one-loop and two-loop form factors that are relevant for the annihilation of dark matter into colored Standard Model particles through a colored scalar mediator, $\phi$ in the triplet representation. The annihilation of dark matter into gluons is a leading-order one-loop process, while that into quarks involves two-loop Feynman diagrams at leading order. The one-loop form factor for the gluon channel has been evaluated up to ${\cal O}(\epsilon^4)$, while the two-loop form factor for the gluon channel is evaluated up to ${\cal O}(\epsilon)$. For the quark channel, the two-loop form factor is evaluated up to ${\cal O}(\epsilon^0)$. The results for the form factors are obtained for both massive and massless colored scalar particles $\phi$ in the loop. 
The analytic resuts for form factors can be combined with real emission contributions to obtain full NLO QCD predictions for dark matter annihilation and production cross section.

\section*{Acknowledgement}
WK would like to acknowledge financial support from the Prime Minister's Research Fellows (PMRF) Scheme for this work. We would like to thank Dr. Narayan Rana for his useful insights and discussions. 

\appendix
\section{Appendix}
\label{app:Appendix}
\subsection{Two-loop form factor for $ g g $ channel with a massless mediator }
\label{sec:ggchannelm0}
In case of $m_\phi=0$, the renormalized amplitude of Eq. \ref{eqn:ggrenorm} becomes,
\begin{equation}
   \mathcal{A}(\alpha_s,\lambda_d,\mu)=Z_g \mathcal{A}^{0}(\alpha_s^{0},\lambda_d^{0}) 
\end{equation}
The renormalization procedure is similar to the massive mediator case except that mass renormalization counter-term is absent and no wavefunction renormalization of external gluon is required.
The universal infrared divergent structure at NLO in this case is given by,
\begin{eqnarray}
    M_{\rm IR}|_{m_\phi = 0} &=& -\dfrac{e^{-\gamma_{E} \epsilon}}{\Gamma(1-\epsilon)}\Big(\dfrac{\beta_0'}{\epsilon}+\dfrac{2N_c}{\epsilon^2}\Big)\mathcal{M}^{\rm 1L,0}_{gg}|_{m_\phi = 0}
    \end{eqnarray}
where,
\begin{equation}
\beta_0'=\dfrac{11}{3}N_c-\dfrac{2}{3}N_f-\dfrac{1}{6}N_\phi
\end{equation}
Note that $\beta_0'$ is modified $\beta_0$ function in the presence of massless colored scalar mediators.
Here $N_\phi$ is the number of colored scalar mediators.

The ultraviolet divergences for the massless mediator will have the structure,
\begin{eqnarray}
    M_{\rm UV}|_{m_\phi = 0} &=& -\Big(\dfrac{-s}{\mu^2}\Big)^{-\epsilon}\Big(\ \delta Z_{\alpha_s}  \  + \ \delta Z_{\lambda} \Big) \ \mathcal{M}^{\rm 1L,0}_{gg}|_{m_\phi = 0}
\end{eqnarray}
with,
\begin{eqnarray}
    \delta Z_{\alpha_s} &=& - \Big(\dfrac{\beta_0'}{\epsilon}\Big);~~
    \delta Z_{\lambda} = - \dfrac{3 C_F}{\epsilon}.
\end{eqnarray}

After renormalization and IR subtraction, the finite part becomes,
\begin{equation}
\mathcal{M}^0_{\rm fin}|_{m_\phi = 0}=  \dfrac{1}{s} \Big( -26~N_c + 14~\dfrac{1}{N_c} \Big)
\end{equation}
There are only 5 master integrals relevant at two-loop order. The analytical results for these are available in Appendix A of Ref.~\citep{Gehrmann:2010ue}.


\subsection{Two-loop form factor for $ q \bar{q } $ channel with a massless mediator }
\label{sec:qqchannelm0}

When $m_\phi=0$, the result depends only of one dimensionless parameter $y$ given by,
\begin{equation}
\dfrac{s}{m_q^2} = \dfrac{-(1-y)^2}{y}.
\label{eq:qqmasslessParametrization}
\end{equation}
The leading order two-loop form factor in this case is give by,
{\allowdisplaybreaks
\begin{equation}
\begin{aligned}
\mathcal{F}_{q\bar{q}}^{\rm 2L}|_{m_\phi = 0}= & \Big(\dfrac{\alpha_s}{4 \pi}\Big)^2~\dfrac{i\lambda_d~C_F~m_q (1-y)}{3 s (y+1)^3} \biggl[
   -12 y^2 \,\mathrm{Li}_2(y)
   + 3 \left(3 y^2 - 6 y - 1\right) \log^2(y) \\
& + 24 \left(y^2 - 1\right) \log^2(1-y)
   + 72 y \,\mathrm{Li}_2(y) 
   - 12 \,\mathrm{Li}_2(y) \\
& + 4 \pi^2 y^2 
   - 24 y^2 \log^2(1-y) 
   - 6 y^2 \log^2(y) 
   + 48 y^2 \log(1-y) 
   - 24 y^2 \log(y) \\
& - 24 \pi^2 y 
   + 24 \log^2(1-y) 
   + 6 \log^2(y) 
   - 48 \log(1-y) 
   + 24 \log(y) 
   + 4 \pi^2 
\biggr]
\end{aligned}
\end{equation}}
The above result is based on analytical results for the master integrals obtained from $l$ and $m$ integral families discussed in Ref.~\citep{Anastasiou:2020qzk}.

\newpage

\bibliographystyle{JHEP}
\bibliography{bibrefer} 
\end{document}